\documentstyle[12pt]{article}
   \font\tenmsb=msbm10 scaled\magstep 1
   \font\sevenmsb=msbm7 scaled \magstep 1
   \font\faivemsb=msbm5 scaled \magstep 1
\newfam\msbfam
      \textfont\msbfam=\tenmsb
      \scriptfont\msbfam=\sevenmsb
      \scriptscriptfont\msbfam=\faivemsb
\def\Bbb#1{{\fam\msbfam #1}}
\font\tengothic=eufm10 scaled\magstep 1
\font\sevengothic=eufm7 scaled\magstep 1
\newfam\gothicfam
      \textfont\gothicfam=\tengothic
      \scriptfont\gothicfam=\sevengothic

\newcommand{\sr}{\stackrel{\rightarrow}{r}}
\newcommand{\sk}{\stackrel{\rightarrow}{k}}
\newcommand{\bt}{\beta}
\newcommand{\al}{\alpha}
\newcommand{\om}{\omega}
\newcommand{\sgm}{\sigma}
\newcommand{\ra}{\rightarrow}
\newcommand{\ga}{\gamma}
\newcommand{\be}{\begin{equation}}
\newcommand{\ee}{\end{equation}}
\newcommand{\dgr}{\dagger}
\newcommand{\Om}{\Omega}
\newcommand{\ep}{\varepsilon}
\newcommand{\vp}{\varphi}
\newcommand{\lgl}{\langle}
\newcommand{\rgl}{\rangle}
\newcommand{\prt}{\partial}

\begin{document}
\begin{flushright}
{\large ""Š 539.12.01}
\end{flushright}

\vspace{1cm}

\begin{center}
{\Large{\bf Thermodynamics of Strong Interactions} \\ [5mm]
V.I.Yukalov and E.P.Yukalova} \\ [3mm]
{\it Joint Institute for Nuclear Research, Dubna, Russia \\ 
and \\ 
Queen's University, Kingston, Canada}
\end{center}

\vspace{1cm}

The state of art in studying thermodynamic properties of hot and dense
nuclear matter is reviewed with the special emphasis on the
confinement--deconfinement transition between hadron matter and quark--gluon
plasma. The most popular models used for describing deconfinement are
analysed, including statistical bootstrap models, pure phase models, the
model of clustered quarks, and the string--flip potential model. Predictions
of these models are compared with the lattice numerical simulations. It is
concluded that precursor fluctuation effects must be taken into account
in order to get a realistic description of deconfinement transition. The
existence of precursor fluctuations is in line with the dynamical confinement
scenario and suggests that deconfinement cannot be considered as a transition
between pure hadron and quark--gluon phases. All this supports the concept
of cluster coexistence advocated by the authors of this review: Quark--gluon
plasma and hadron clusters are different quantum states of the same system,
so that any statistical model pretending to treat nuclear matter under extreme
conditions must incorporate into itself the probability of these different
channels. The ways of constructing statistical models with plasma--cluster
coexistence are discussed and thermodynamic properties of such models are
analysed.

\section{Introduction}

One of the most intriguing problems of high energy physics is the possibility
of transforming the nuclear matter composed of hadrons into the phase
consisting of their fundamental constituents, quarks and gluons. This phase,
because of the  apparent analogy with the electron--ion plasma, is called the
quark--gluon plasma. The transformation of the hadron matter into the
quark--gluon plasma is named deconfinement; the inverse process, respectively,
being called confinement. The deconfinement transition is somewhat similar to
the ionization of atoms. The literature devoted to this phenomenon is so
numerous that, not to overload the list of references, we shall cite here
mainly review papers, when these are available, and in which the reader can
find thousands of references to original works. The specific feature of the
present review is that we concentrate on the statistical models of strongly
interacting systems under extreme conditions, when non--hadronic degrees of
freedom become important and deconfinement occurs.

The possibility that quark degrees of freedom can come into play in the
process of relativistic nuclear collisions at already achieved accelerator
energies was advances by A.M.Baldin [1] who predicted and explained the
commulative effect as a manifestation of the formation in the colliding
nuclei of multiquark droplets.

In order that quark--gluon degrees of freedom would be essential, special
conditions are necessary, like high temperature or density. From quantum
chromodynamics it is known that at asymptotically high temperature quarks
and gluons are really deconfined forming the quark--gluon plasma [2-6]. We
also know that at zero temperature and at the normal density of nuclear
matter there is complete confinement so that only hadrons exist. But where
is the intermediate region in which quark--gluon degrees of freedom become
relevant? The characteristic parameters for this region can be estimated as
follows.

Nuclear matter in the normal state has the baryon density
$\;n_{0B} = 0.167/fm^3\;$. The corresponding normal quark density is
$\;\rho_0\equiv 3n_{0B}=0.5/fm^3\;$. The characteristic quark interaction
energy in the normal state can be presented as $\;E_0 =\hbar /\tau_0\;$, with
the interaction time $\;\tau_0 =a_0/c\;$, where $\;a_0 =\rho_0^{-1/3}\;$
is the mean interquark distance in the normal state. Accepting the system
of units in which $\;\hbar = c=1\;$, with the conversion constant
$\;\hbar c = 197.327\;MeV \; fm \;$, we have
$$ E_0 =\rho_0^{1/3} =157\;MeV . $$
Note an interesting relation between this interaction energy and the
characteristic baryon energy density $\;\varepsilon_B \equiv m_Nn_{0B}\;$,
in which $\; m_N\equiv\frac{1}{2}(m_p + m_n) =939\;MeV\;$ is the average
nucleon mass. Since $\;\ep_B=157\;MeV/fm^3\;$, therefore $\;E_0=\ep_B\;fm^3\;$.
Thus, the normal nuclear matter should start to decompose
being heated up to the temperature equal to the quark interaction energy
$\;E_0\;$, that is up to $\;\Theta_c \approx 160\;MeV\;$.

If one wishes to destroy the hadron matter by compression, one has to reach
a density of about the density of quarks inside a nucleon. This
characteristic density is $\;\rho_c\equiv 3/v_N\;$, where
$\;v_N\equiv (4\pi /3)r_N^3\;$ is the nucleon volume. Taking for the nucleon
radius $\;r_N=0.9\;fm\;$, we get $\;v_N=3\;fm^3\;$. From here
$$ \rho_c =2\rho_0 =1/fm^3 . $$
This tells us that the hadron matter should start desintegrating being
compressed up to the density $\;\rho_c\approx 2\rho_0\;$.

As we see, the expected critical temperature and density are fairly low. Such
conditions certainly existed in the early Universe about $\;10^{-5}\div
10^{-4}sec\;$ after the Big Bang and are likely to exist in the interior of
neutron stars [7].

More important is the common belief that such conditions can be created in the
process of relativistic heavy ion collisions, even with existing accelerators,
thus opening the path for experimental observation of the quark--gluon plasma
in the laboratory [8-14]. Analogous relativistic nuclear collisions can
also be studied in cosmic ray experiments [8]. The density of matter inside
a fireball formed by two collided ions can reach $\;10\rho_0\;$.

All models of the formation of the quark--gluon plasma in nuclear collisions
require the information about its rate of thermalization. Does the matter
inside a fireball thermalize sufficiently fast, so that a thermodynamic
description makes sense? For this, the thermalization time must be shorter
than the fireball lifetime $\;\tau_f\sim 10^{-22}sec\;$. The thermalization
time, or the time of local equilibration, $\;\tau_{loc}\;$, can be estimated
as follows [13-15]. The local equilibrium time writes as $\;\tau_{loc} =
\lambda /c\;$, in which $\;\lambda =(\rho\sigma )^{-1}\;$ is the mean free
path of a particle in a medium; $\;\rho =a^{-3}\;$, the density of matter;
$\;a\;$, average interparticle distance; $\;\sigma \sim b^2\;$, crossection;
$\;b\;$, interaction radius. Accepting the values $\;a\sim b\sim 1fm\;$
typical of nuclear matter, we have $\;\lambda \sim 1fm\;$ and
$\;\tau_{loc}\sim 10^{-23}sec\;$. Because of the inequality
$\;\tau_{loc}\ll\tau_f\;$, the thermalization inside a fireball is
likely to be reached.

The possibility of speaking about the thermodynamics of strong interactions,
separately from electromagnetic and weak interactions, is based on the fact
that it is just this type of interaction which in many cases plays the
dominant role. Really, the dimensionless coupling constant of strong
interactions $\;\alpha_s\approx 1\;$ is much larger than the coupling
constants of electromagnetic interactions,
$\;\alpha_e \approx 1/137 \sim 10^{-2}\;$, and of weak interactions,
$\;\alpha_w\sim 10^{-5}\;$, to say nothing of the coupling constant of
gravitational interactions, $\;\alpha_g\sim10^{-12}\;$. The corresponding
interaction times, at the energy $\;1\;GeV\;$ characteristic of high--energy
physics, are $\;\tau_s\sim 10^{-24}sec\;$ for strong interactions, 
$\;\tau_e\sim 10^{-21}sec\;$ for electromagnetic interactions, and
$\;\tau_w\sim 10^{-10}sec\;$ for weak interactions. Therefore, during
the lifetime of a fireball $\;\tau_f\sim 10^{-22}sec\;$ electromagnetic
and weak interactions do not play any role. In the interval of time
$\;\tau_{loc} < t<\tau_f\;$ inside a fireball there may exist an equilibrium
state of strongly interacting particles.

The sole consistent way of calculating thermodynamic characteristics in the
frame of quantum chromodynamics is perturbation theory, which is quite
similar to that of quantum electrodynamics [16,17]. However, the effective
coupling parameter of strong interactions becomes small only at
asymptotically high temperatures. In the most interesting region of
temperatures around $\;\Theta_c\approx 160\;MeV\;$, where deconfinement occurs,
the coupling parameter is large and perturbation theory does not work. For
describing the whole range of thermodynamic variables several statistical
models have been suggested .

\section{Statistical Bootstrap Models}

The usefulness of applying statistical methods for considering the heated
and compressed nuclear matter has been understood long time ago. Let us
mention, e.g.,Fermi [18].

The first statistical model of nuclear matter under extreme conditions,
such as being realized inside fireballs, has been proposed by Hagedorn [19]
(see also [20]) and called the statistical bootstrap model. In this approach
it is assumed that, at zero baryon density $\;n_B =0\;$, various hadrons can
be generated from vacuum with the mass  distribution
$$ \rho (m) =\rho_{dis}(m) +\rho_{con}(m) ; \qquad m\in [0,\infty ) , $$
in which the first and the second terms correspond to discrete and to
continuous mass spectra respectively,
$$ \rho_{dis}(m) =\sum_{i}\zeta_i\delta (m-m_i)
\left [ 1 -\Theta ( m - m_0 ) \right ] , $$
$$ \rho_{con}(m) = \Theta (m-m_0)
\frac{a_0}{m^{5/2}}\exp\left (\frac{m}{\Theta_0} \right ) , $$
where $\;\Theta (\cdot )\;$ is the unit--step function; $\;\zeta_i\;$, a
degeneracy number for spin--isospin states; and the parameters are
$$ m_0 =1000\;MeV, \quad a_0 = 6.5\times 10^3\; MeV^{3/2} , \quad
\Theta_0 =160\;MeV . $$
The pressure for the ideal hadron gas is written in the classical Boltzmann
approximation,
$$ p =\Theta\int \rho (m)\exp\left ( -\beta\sqrt{k^2+m^2}\right )
\frac{d\stackrel{\rightarrow}{k}}{(2\pi )^3} dm , $$
where $\;\Theta\;$ is temperature in energy units, and $\;\beta\;$ is
inverse temperature, $\;\beta\Theta\equiv 1\;$.

As can be easily checked, the pressure in this model diverges for all
$\;\Theta \geq \Theta_0 \;$. From here it was concluded [19] that
$\;\Theta_0\;$ is the limiting temperature of the Universe. The concept
of the existence of a maximal temperature of the Universe is, of course,
quite artificial, therefore another interpretation of this divergency of
pressure has been proposed  [20] treating $\;\Theta_0\;$ as the deconfinement
temperature.

After Collins and Perry [21], the geometrical scenario of the deconfinement
became popular [22]. According to this, the increase of temperature or
baryon density leads to the rising number of hadrons. The latter are
assumed to have finite volumes [23]. When the number of hadrons becomes so
high that their close packing occurs, then they fuse into one gigantic
hadron occupying the whole system. The Hagedorn temperature $\;\Theta_0\;$
is interpreted as the fusion temperature, and the gigantic hadron cluster is
identified with the system in the quark--gluon--plasma state. This geometric
scenario reminds the percolation transition [24].

Following the geometric interpretation, the bootstrap model was modified [20]
invoking the excluded--volume approximation. In this, one considers $\;N\;$
particles having the volumes $\;v_1,v_2,\ldots ,v_N\;$ as moving in the free
volume
$$ V_N \equiv V -\sum_{j=1}^{N}v_j , $$
where $\;V\;$ is the total volume of the system. The pressure in the
excluded--volume approximation becomes
$$ p =\frac{\Theta}{V}\ln \left \{ \sum_{N=1}^{\infty}
\frac{1}{N!}\Theta (V_N) 
\left [ V_N\int \rho (m) \exp \left ( -\beta\sqrt{k^2+m^2} \right )
\frac{d\stackrel{\rightarrow}{k}}{(2\pi )^3}dm\right ]^N\right \} , $$
where again the Boltzmann approximation is also used. The mass distribution
for discrete spectrum is taken in the same form as above, and for continuous
spectrum it is slightly modified as
$$ \rho_{con}(m) =\Theta (m-m_0)
\frac{a_0}{m^\alpha}\exp\left (\frac{m}{\Theta_0}\right ) , $$
with $\;\frac{3}{2} <\alpha <\frac{7}{2}\;$. Now, the pressure is everywhere
finite and positive becoming zero at the same temperature
$\;\Theta_d\approx \Theta_0\;$. The temperature $\;\Theta_d\;$, where
$\;p(\Theta_d)=0\;$ is interpreted as the temperature of hadron fusion
into a gigantic cluster. However, the thermodynamics of the system at
$\;\Theta\rightarrow \infty\;$ has nothing to do with that of the ideal
quark--gluon plasma.

To overcome the latter deficiency of the model, it has been argued that
taking into account hadron compression can save the situation. This can
be done [25] by complicating the mass distribution writing its continuous
part as
$$ \rho_{con} (m,v) =\Theta (m-b_0v-m_0)\Theta (v-v_0)a_0(m-b_0v)^\alpha
\times  $$
$$ \times v^\gamma\exp \left \{\frac{4}{3}
\left (\sigma_0v\right)^{1/4} \left (m -b_0v\right )^{3/4}\right \} , $$
which contains now seven fitting parameters: $\;m_0,a_0,\alpha,b_0,v_0,
\gamma,\sigma_0\;$.

Now, in accordance with the geometrical scenario, the number of hadrons at
low temperature $\;\Theta < \Theta_d\;$ is proportional to the free volume,
and at high temperatures $\;\Theta \rightarrow \infty \;$ this number
tends to one symbolizing the formation of a gigantic cluster. However,
at the 1--st order transition temperature $\;\Theta_d\;$ the number of
hadrons $\;N(\Theta_d,V)\;$ diverges for any finite volume $\;V\;$, which
is unreasonable.

The bootstrap models, in addition to the arbitrariness in postulating the mass
distribution, contain internal deficiencies leading to the existence of
instabilities contradicting the necessary stability conditions for
statistical systems [26,27].

\section{Pure Phase Models}

An evident idea would be to follow the standard Gibbs prescription for
considering phase transitions between two phases. Treating the deconfinement
as such a phase transition, one assumes the existence of two types of pure 
phases. At low temperatures and baryon densities this is a pure hadron phase
with features typical of the normal nuclear matter [28], and at high
temperature or baryon density this is the quark--gluon phase described by
perturbative QCD [17]. Let $\;\Omega_1\;$ be the grand potential of the
quark--gluon plasma, and $\;\Omega_2\;$, that of the hadron matter. According
to the Gibbs rule, a phase transition occurs when $\;\Omega_1(\Theta ,\mu_B)
=\Omega_2(\Theta ,\mu_B)\;$, where $\;\mu_B\;$ is the baryon chemical
potential. This equality gives a transition line
$\;\Theta_d =\Theta_d(\mu_B)\;$. Expressing here the baryon potential
$\;\mu_B=\mu_B(n_B)\;$ through the baryon density, we may write
$\;\Theta_d=\Theta_d(n_B)\;$.

The possibility of using perturbation theory for the high--temperature
quark--gluon plasma is based on the property of asymptotic freedom. According
to this property, the running coupling constant
$$\alpha_s(q) \simeq
\frac{6\pi}{(11N_c-2N_f)\ln (q/\Lambda )}\left [ 1 -
\frac{51\ln\ln (q/\Lambda )}{121 \ln (q/\Lambda )}\right ] , $$
in which $\;N_c\;$ and $\;N_f\;$ are the number of quark colours and flavours,
respectively, $\;\Lambda\approx 200\;MeV\;$ is a scale parameter, and $\;q\;$
is momentum, tends to zero as $\;q\rightarrow \infty\;$. In the integral over
momenta, defining the grand potential, the main contribution, when
$\;\Theta\rightarrow\infty\;$, comes from $\;q\approx \Theta\;$. Hence, it is
possible to get an expansion in powers of $\;\alpha_s(\Theta )\equiv
g^2(\Theta )/4\pi\;$ with the effective coupling parameter
$$ g^2(\Theta ) \simeq \frac{24\pi^2}{(11N_c-2N_f)\ln (\Theta/\Lambda)} . $$
As a result of this expansion [29], neglecting quark masses, one has for
the grand potential
$$ \frac{\Omega_1}{V} = -A\Theta^4 +B , $$
in which a nonperturbative term $\;B\;$ is included and the notation
$$ A\equiv A_0 + A_2g^2 +A_3g^3 +A_4g^4\ln g $$
is used, where
$$ A_0 =\frac{\pi^2}{45} \left [ N_c^2 - 1 +\frac{7}{4}N_cN_f +
15N_c\sum_{f}\frac{\mu_f^2}{2\pi^2\Theta^2}
\left ( 1 + \frac{\mu_f^2}{2\pi^2\Theta^2}\right )\right ] , $$
$$ A_2 =-\frac{N_c^2-1}{144} \left [ N_c +\frac{5}{4}N_f +
9\sum_{f}\frac{\mu_f^2}{2\pi^2\Theta^2}
\left ( 1 + \frac{\mu_f^2}{2\pi^2\Theta^2}\right )\right ] , $$
$$ A_3 =\frac{N_c^2-1}{12\pi}\left (\frac{1}{3} N_c +\frac{1}{6}N_f +
\frac{1}{3}\sum_{f}\frac{\mu_f^2}{2\pi^2\Theta^2}\right )^{3/2} , $$
$$ A_4 =\frac{N_c^2-1}{16\pi^2}N_c\left (\frac{1}{3} N_c +\frac{1}{6}N_f +
\sum_{f}\frac{\mu_f^2}{2\pi^2\Theta^2}\right ) ; $$
$\;\mu_f\;$ being the chemical potential of an $\;f\;$--flavour quark. Note
that the term $\;O(g^6)\;$ cannot be calculated by perturbation theory
because of unrenormalizable infrared divergences.

Take into account that the number of quark colours is $\;N_c=3\;$, and
consider for simplicity the case of zero baryon density $\;n_B=0\;$, so that
$\;\mu_f=0\;$. Then the pressure of the quark--gluon plasma is
$$ p_1\equiv -\frac{\Omega_1}{V} =A\Theta^4 - B , $$
with the expansion coefficients of $\;A\;$ being
$$ A_0 =\frac{8\pi^2}{45}\left ( 1 +\frac{21}{32}N_f\right ) , \qquad
A_2 =-\frac{1}{6}\left ( 1 + \frac{5}{12}N_f\right ) , $$
$$ A_3 =\frac{2}{3\pi}\left ( 1 +\frac{1}{6}N_f\right )^{3/2} , \qquad
A_4 =\frac{3}{2\pi^2}\left ( 1 +\frac{1}{6}N_f\right ) . $$

The low--temperature hadron phase is often modeled [4,7,30] by a gas
of massless noninteracting pions, which for the grand potential yields
$$ \frac{\Omega_2}{V} = -\frac{\pi^2}{30}\Theta^4 . $$
The corresponding pressure is
$$ p_2 \equiv -\frac{\Omega_2}{V} =\frac{\pi^2}{30}\Theta^4 . $$

Equating $\;\Omega_1\;$ with $\;\Omega_2\;$, or $\;p_1\;$ with $\;p_2\;$,
one obtains the deconfinement temperature
$$ \Theta_d =\gamma B^{1/4} ; \qquad \gamma\equiv\left ( A -
\frac{\pi^2}{30}\right )^{-1/4} . $$
To check the phase transition order, let us find the latent heat at
the transition temperature.

Define the energy density
$$ \varepsilon = s\Theta - p + \mu_Bn_B $$
and the entropy density
$$ s = -\frac{\partial}{\partial \Theta}\left (\frac{\Omega}{V}\right )
=\frac{\partial p}{\partial \Theta} . $$
The latent heat density is
$$ \Delta\varepsilon_d\equiv\varepsilon_1 -\varepsilon_2 =\Theta_d\Delta s_d
\qquad (\Theta =\Theta_d) , $$
where $\;\Delta s_d\equiv s_1 -s_2\;$ is the entropy density jump at
$\;\Theta =\Theta_d\;$.

For the considered case we have
$$ s_1 =(4A +C)\Theta^3 , \qquad s_2 =\frac{2\pi^2}{15}\Theta^3 , $$
with the notation
$$ C\equiv \Theta\frac{\partial A}{\partial \Theta} = -
\frac{33-2N_f}{24\pi^2}g^4 \left [ A_2 +\frac{3}{2}A_3g +
\frac{1}{2}A_4\left ( 1 +4\ln g \right ) g^2\right ] , $$
where the equation
$$ \frac{\partial g^2}{\partial \Theta} = -
\frac{33-2N_f}{24\pi^2\Theta}g^4 $$
is taken into account. The energy densities are
$$ \varepsilon_1 = (3A + C)\Theta^4 + B , \qquad \varepsilon_2 =
\frac{\pi^2}{10}\Theta^4 . $$
Thus, for the latent heat one gets
$$ \Delta\varepsilon_d = 4B\left ( 1 + \frac{C}{4}\gamma^4\right ) . $$

The nonperturbative term $\;B\;$ is usually treated as the bag constant,
with $\;B^{1/4}\;$ ranging in the interval $\;(150\div 300)\;MeV\;$. For
estimates, we may accept the value $\;B^{1/4} =225\;MeV\;$ from the middle of
this interval.

One often assumes that the quark--gluon plasma is an ideal gas of free
quarks and gluons, that is $\;g=0\;$. If this is so, then for $\;N_f=2\;$
we have $\;\gamma=0.72\;$. The deconfinement occurs at $\;\Theta_d=162\;MeV\;$
being a first--order transition with the latent heat
$\;\Delta\varepsilon_d=4B\approx 1\;GeV/fm^3\;$. The found deconfinement
temperature $\;\Theta_d\;$ almost coincides with the characteristic energy
$\;E_0=157\;MeV\;$ discussed in Introduction. This picture would seem quite
reasonable if it would not be absolutely wrong. Really, the effective
coupling $\;g(\Theta )\rightarrow\infty\;$ for temperatures $\;\Theta\;$
close to the scale parameter $\;\Lambda\approx 200\;MeV\;$. Therefore, the
assumption that, in the vicinity of the deconfinement point, the quark--gluon
plasma is an ideal gas of quarks and gluons is senseless. This would happen
only at temperatures at which $\;g^2(\Theta ) \ll 1\;$, that is when
$$ \Theta \gg \Lambda\exp\left ( \frac{24\pi^2}{33-2N_f}\right ) . $$
The latter inequality becomes valid only at very high temperatures
$\;\Theta \gg 10^6\;MeV\;$.

In this way, nonperturbative effects around the deconfinement transition are
very strong, and it is not correct to try taking account of them by the
simple addition to the grand potential of a term $\;B\;$.

Some nonperturbative effects can be included into consideration by resorting
to the effective--spectrum approximation [31,32], when one postulates for
the spectra of quarks and gluons the form
$\;\varepsilon_i(k) =\sqrt{k^2+m_i^2}+U_i\;$, in which $\;k\;$ is the modulus
of momentum; $\;m_i\;$, a mass; $\;U_i\;$, an effective mean field; and
$\;i=q,g\;$ enumerates quarks and gluons. This approximation yields
the results similar to the ideal gas picture: The deconfinement is a
first--order transition occurring at $\;\Theta\approx 160\;MeV\;$.

The reason why the effective--spectrum and ideal--gas approximations are
close to each other can be understood as follows. The effective--spectrum
approximation may be interpreted as a result of a renormalization of
perturbative series. As an example, we may use the self--similar
renormalization [33-36] differing from other resummation techniques by the
possibility of checking its range of applicability at each step. Consider
the coefficient $\;A\;$ in the grand potential $\;\Omega_1\;$ of
the quark--gluon plasma as an effective limit of the sequence
$\;\{ f_k(g)\}\;$ with the initial approximation $\;f_0(g)=A_0 +A_2g^2\;$,
the first approximation $\;f_1(g)=f_0(g)+A_3g^3\;$, and so on. The simplest
variant of the self--similar renormalization [33-36] gives the renormalized
coefficient
$$ A^* = A_0 +\frac{4A_2^3g^2}{(A_3g-2A_2)^2} . $$
This quantity, as $\;A_2 < 0\;$, is finite for all $\;g\;$ including
$\;g\rightarrow\infty\;$. In the whole diapason of $\;g\in [0,\infty )\;$
the  deconfinement temperature does not change much: for $\;g=0\;$, with
$\;N_f=2\;$, it is $\;\Theta_d=162\;MeV\;$, while for $\;g\rightarrow\infty\;$
it is $\;\Theta_d=176\;MeV\;$. However, the renormalized value $\;A^*\;$ is
truthful only when the self--similar renormalization is stable [37-40].
For this we need that the corresponding mapping multiplier $\;M_1(g)\;$
and the Lyapunov exponent $\;\Lambda_1(g)\;$ would satisfy the stability
conditions: $\;|M_1(g)| < 1\;$ and $\;\Lambda_1(g) < 0\;$. In the considered
case
$$ M_1(g) = 1 + \Lambda_1(g) , \qquad \Lambda_1(g) =\frac{3A_3}{2A_2}g . $$
The Lyapunov exponent, since $\;A_2<0\;$ and $\;A_3>0\;$, is always negative.
The condition $\;M_1(g)\;$ holds only for $\;g<4|A_2|/3A_3\;$, that is for
$\;g \leq 1\;$.

Thus, the renormalization of the grand potential, starting from the
perturbative expression, does not essentially change the results. And it is
clear why: Really, the ideal--gas picture, which makes the basis of the
perturbative expansion, contains no information on bound states that should
appear as the coupling constant $\;g\;$ increases. The situation with the
quark--gluon plasma is somewhat similar to that of the electron--ion plasma
in which there can exist bound as well as free electron states [41].

The value of the deconfinement temperature obtained in pure--phase models
can be quite reasonable, in the same way as the simple estimate of
Introduction is such. However, it would be hard to believe that these
models can correctly describe the character of the deconfinement transition
and the behaviour of thermodynamic functions.

\section{Lattice Numerical Simulation}

The idea of using a discrete space--time lattice to regularize quantum
field theories opened the entire repertory of statistical physics for the
analysis of nonperturbative properties of these theories. The application
of Monte Carlo simulation techniques turned out to be a powerful approach
allowing to perform a quantitative study of nonperturbative aspects of
quantum chromodynamics. The lattice reformulation of $\;QCD\;$ has been
described in several surveys (e.g.[4,29,42]), therefore below we only
slightly touch the principal points of this approach. We shall mainly discuss
the predictions of the lattice $\;QCD\;$ and its simplified variants for the
deconfinement transition.

The first step towards the finite temperature study of $\;QCD\;$ consists of
introducing the imaginary time $\;t=-i\tau\;$ and of rewriting the partition
function as a path integral of the exponential of the Euclidean Lagrangian
density over all fields in the problem. The second step defines the cubic
four--dimensional lattice with the sites $\;x=\{\stackrel{\rightarrow}{x},
\tau\}\;$ where $\;\stackrel{\rightarrow}{x}\in \Bbb{Z}_3\;$ is a real--space
lattice vector and $\;\tau\in\Bbb{Z}_1\;$ is conventionally called the
temporal variable. The lattice spacings in the spatial and temporal
directions are denoted by $\;a_\sigma\;$ and $\;a_\tau\;$, respectively.
If $\;N_\sigma\;$ and $\;N_\tau\;$ are the number of sites in the
corresponding directions then the volume $\;V\;$ and the temperature
$\;\Theta\;$ of the system are given by $\;V\equiv (N_\sigma a_\sigma)^3\;$
and $\;\beta\equiv N_\tau a_\tau\;$, where $\;\beta\equiv\Theta^{-1}\;$.

A gauge--invariant theory on the lattice is usually formulated in terms
of link variables $\;U_x^\mu\;$ and site variables $\;\psi(x)\;$ and
$\;\bar\psi(x)\;$. The link variable $\;U_x^\mu\;$ is associated with a link
leaving site $\;x\;$ in a direction $\;\mu=1,2,3,4\;$ and it is a matrix
$\;U_x^\mu\in SU(N_c)\;$ in the space of colour indices, $\;N_c\;$ being the
number of colours. The site variables $\;\psi(x)\;$ and $\;\bar\psi(x)\;$
are associated with each site $\;x\;$ of the lattice and they carry colour,
flavour and spin indices. Also, $\;\psi(x)\;$ and $\;\bar\psi(x)\;$,
representing fermion fields, are treated as Grassman variables. The link
and the site variables satisfy the periodicity conditions
$$ U^\mu_{\stackrel{\rightarrow}{x},0} =
U^\mu_{\stackrel{\rightarrow}{x},\beta} , \qquad
\psi(\stackrel{\rightarrow}{x},0) =
- \psi(\stackrel{\rightarrow}{x},\beta) . $$

Then, in terms of these variables, the partition function is written in
the form of a path integral
$$ Z =\int \prod_{x}\prod_{\mu}\prod_{x'}{\rm e}^{-S}dU_x^\mu
d\psi(x')d\bar\psi(x') , $$
in which the action
$$ S = S(U_x^\mu ,\psi(x')\bar\psi(x')) = S_G + S_F $$
consists of a gauge action $\;S_G\;$ and a fermionic action $\;S_F\;$. The
partition function and other thermodynamic functions are calculated by using
the Monte Carlo procedure on a finite lattice of $\;N_\sigma^3\times N_\tau\;$
sites, the maximal number of sites in each direction being around
$\;N_\sigma=24\;$ and $\;N_\tau=24\;$.

The deconfinement transition has been studied, in the frame of the lattice
$\;QCD\;$ or its pure gauge variants, by many authors. Here we cite only some
review--type papers [43-52]. Note, that, in addition to the confinement
transition, another transition related to the  spontaneous breaking of
chiral symmetry can occur. The mechanisms leading to these transitions are
seemingly unrelated and it thus has been speculated that $\;QCD\;$ may
undergo two separate phase transitions. However, the Monte Carlo data for
pure gauge models and for the variants with massless quarks suggest that
these two transitions coincide. And if the finite masses of quarks are
taken into account then the chiral symmetry, strictly speaking, happens
only in the limit $\;\Theta\rightarrow\infty\;$. In what follows we shall
speak solely about the deconfinement transition. One of the main discoveries
of the lattice simulation has been the fact that the deconfinement can be
of quite different character for different systems.

\vspace{5mm}

$\;SU(2)\;$ Pure Gauge Model

\vspace{2mm}

The common convention is that the deconfinement for such quarkless models is 
a {\it second--order transition} occurring at $\;\Theta_d\approx 210\;MeV\;$.
The errors in calculating the pressure and energy density, related to the
finiteness of the used lattice, are around $\;5\%\;$ above and $\;30\%\;$
below $\;\Theta_d\;$.

\vspace{5mm}

$\;SU(3)\;$ Pure Gauge Model

\vspace{2mm}

The deconfinement has been found to be a {\it first--order transition}
at $\;\Theta_d\approx 225\;MeV\;$ with the latent heat
$\;\Delta\varepsilon_d\approx\frac{1}{4}\varepsilon_1(\Theta_d)\;$. The
errors of calculating the pressure, energy density and entropy are about
$\;10\%\;$ above and $\;30\%\;$ below $\;\Theta_d\;$.

\vspace{5mm}

$\;SU(3)\;$ Model with Quarks

\vspace{2mm}

In the presence of dynamical fermions the situation is by far more
complicated as it is difficult to perform high statistics analysis of full
$\;QCD\;$ on large lattices that would allow a detailed finite size study as
it has now been done in the pure gauge sector. Monte Carlo simulations for
full $\;QCD\;$ with dynamical quarks have by now been performed only for
the case of zero baryon density. The transition has been found to be rather
sensitive to the choice of quark masses and the number of flavours. It seems
that for $\;N_f>4\;$ the transition is first order for all quark masses.
For $\;N_f\leq 4\;$ the situation is still to some extent uncertain.
Nonetheless, there are strong indications that for physical quark masses
and $\;N_f=3\;$ the transition is likely to be a continuous crossover,
occurring at $\;\Theta_d\approx 150\;MeV\;$. However, one should once again
stress the uncertainties in the determination of the order of the transition
in Monte Carlo simulations, with quarks, even at zero baryon density.

What has been found in the lattice simulations with certainty [52,53] is
that nonperturbative effects are quite strong around the deconfinement
transition persisting till about $\;2\Theta_d\;$. Also, as is seen, no
statistical bootstrap or pure phase model is able to describe the variety of
different transition orders discovered in the lattice simulations, since
these models always predict a sharp first--order transition.

\section{Dynamical Confinement Scenario}

Correlation functions and related susceptibilities are one of the key tools
used for investigating phase transitions [54]. This concerns as well
$\;QCD\;$ correlation functions [55]. The latter have been intensively
studied in lattice numerical simulations [56-60]. The results of these
studies revealed nontrivial effects in the high temperature phase connected
with strong quark--antiquark correlations, which was interpreted [56-60]
as the existence of hadronic modes, even at high $\;\Theta >\Theta_d\;$.
De Tar [56] proposed that the high temperature phase might be dynamically
confined, in the sense that the long range fluctuations are colour singlet
modes, and that the poles and cuts in the linear response functions of the
hadronic phase go over smoothly into those of the high temperature one.
This scenario of dynamical confinement presupposes that, actually, there is
no transition at all, and that there is only a smooth crossover between
hadronic matter and the quark--gluon plasma.

It is conjectured [56] that the characterization of the plasma as a weakly
interacting gas of quarks and gluons is valid only for short distances and
short time scales of the order $\;1/\Theta\;$, but that at the scales larger
than $\;1/g^2\Theta\;$, where $\;g^2\;$ is the running $\;QCD\;$ coupling,
the plasma exhibits confining features similar to that of the low--temperature
hadronic phase. The confinement scale, for instance, at
$\;\Theta\approx 200\;MeV\;$, is expected to be of order $\;1\;fm\;$ and
$\;10^{-24}sec\;$.

Hatsuda and Kunihiro [61-64] considered the  density--density correlation
functions of quarks for the Nambu--Jona--Lasinio model with a
$\;QCD\;$--motivated effective Lagrangian. They found precursory collective
excitations existing in the high--temperature phase and corresponding to
correlated $\;q\bar q\;$ pairs. This means that meson modes do exist
above as well as below transition temperature. Really, all poles and cuts of
the correlation functions in the momentum--energy representation are the
same at all temperatures. Thus, quarks, antiquarks, and gluons should also
exist in the low--temperature as well as in high--temperature phase. Such
precursor effects are analogous to pretransitional fluctuations in
superconductors [65] or superfluid $\;He^3\;$ [66].

The gradual change of the excitation spectrum from hadronic states to
quarks and gluons, and the survival of hadronic modes above the transition
temperature, have been confirmed for $\;QCD\;$ in the instanton--liquid
approach [67] and in the magnetic--current approximation [68]. In the latter
case, quarks are assumed to interact at high temperature solely through
magnetic current--current interactions, the electric ones being screened.

The following picture [68] may serve as an intuitive illustration of the
dynamical confinement [56-60]: The current--current interactions persist
above $\;\Theta_d\;$ and force any quark--antiquark pair (and, may be, every
three--quark state) to correlate into colour singlets. As the quarks are
moving in the heat bath, the string connecting them for colour neutrality are
constantly breaking and reforming, which can be interpreted as hadrons going
in and out of the heat bath.

If the transition between the hadron matter and the quark--gluon plasma is
a gradual crossover, as follows from the dynamical confinement scenario,
then why in many lattice simulations this transition is found to be of first
order? The answer to this question was suggested
by Kogut et al. [69] explaining the first order of this transition
merely as due to the finite size of the lattice. Calculating thermodynamic
characteristics, such as energy densities, and correlation functions for
several lattices with the sizes $\;8^3\times 4,\; 12^3\times 4,\;
12^3\times 6,\;$ and $\;16^3\times 4\;$, it has been found [69] that the
quark--gluon plasma transition becomes less abrupt as the lattice size
increases and the evidence for a first order transition becomes weaker.
Lengthy runs for $\;N_\sigma = 16\;$ showed no evidence for metastability,
so that Kogut et al. [69] concluded that their results suggest that there is
no abrupt transition at all, but only a smooth crossover phenomenon.

To describe the coexistence of gluons and glueballs above $\;\Theta_d\;$
for a quarkless $\;SU(2)\;$ system, De Grand and De Tar [57,58] proposed to
write the grand potential as the sum
$$ \Omega =\Omega_1 + \Omega_2 , $$
$$ \Omega_1 = -\Theta V \frac{\zeta_1}{2\pi^2}\int\limits_{k_0}^{\infty}
k^2\ln [1 +n_1(\stackrel{\rightarrow}{k})]dk , $$
$$ \Omega_2 = -\Theta V \frac{\zeta_2}{2\pi^2}\int\limits_{0}^{k_0}
k^2\ln [1 +n_2(\stackrel{\rightarrow}{k})]dk , $$
where $\;\Omega_1\;$ corresponds to gluons and $\;\Omega_2\;$, to glueballs;
$\;\zeta_i\;$ is a degeneracy factor; the Bose distribution
$\;n_i(\stackrel{\rightarrow}{k})=
[\exp\{\beta\varepsilon_i(\stackrel{\rightarrow}{k})\}-1]^{-1}$ contains
the spectrum $\varepsilon_i(\stackrel{\rightarrow}{k})$$=\sqrt{k^2+m_i^2}\;$
with the mass $\;m_1=0\;$ for gluons and the mass $\;m_2\approx 1000\;MeV\;$
for glueballs. The cut--off momentum $\;k_0\;$ is postulated, to fit lattice
data, to be $\;k_0\rightarrow\infty\;$ below $\;\Theta_d\;$ and
$\;k_0=2.86\Theta_d[\Theta_d/(\Theta -\Theta_d)]^{0.3}\;$ above
$\;\Theta_d\;$. This phenomenological model describes only the second--order
transition in pure $\;SU(2)\;$ systems. The temperature dependence of the
cut--off momentum $\;k_0\;$ is too arbitrary to permit a straightforward
generalization to systems containing quarks as well as various hadrons.

The main value of the dynamical confinement scenario is the clear
understanding that hadron and quark--gluon degrees of freedom, generally,
coexist.

\section{Clustered Quark Model}

A model in which quarks could coexist with three--quark clusters, that is,
nucleons has been suggested by Clark et al. [70] for zero temperature and
considered by Bi and Shi [71] at finite temperatures. Nucleons and quarks
in this model are intermixed inside the same system. This kind of mixture
should be distinguished from the Gibbs mixture, in which different phases
are separated in space having only a common interphase boundary (see [15]).
The quark--nucleon mixture is rather similar to a binary liquid mixture [72].
A two--component system can, generally, stratify in space if there are no 
chemical reactions between components. The quark--nucleon mixture does not
stratify because of the possibility of formation and decay of nucleon 
clusters, which is taken into account by the relation $\;\mu_N=3\mu_q\;$
between the nucleon, $\;\mu_N\;$, and quark, $\;\mu_q\;$, chemical potentials.
Since it is assumed that inside a nucleon bag and outside it the $\;QCD\;$
vacuum is different, the nonstratified mixture corresponds to nonuniform, or
unhomogeneous, vacuum. In some sense it reminds the twinking vacuum discussed
with respect to the instanton--liquid picture of the $\;QCD\;$ vacuum [29]. 
A nucleon can be interpreted as a droplet of a denser phase inside the 
rarified plasma phase. Nucleons among quarks are alike fog drops in air.
Another analogy is magnetic bubbles in magnets [73].

Clark et al. [70] used the excluded--volume approximation when particles
move not in the whole volume $\;V\;$ but only in the free volume
$\;V_0=V-N_Nv_N\;$, where $\;N_N\;$ is the number of nucleons of the volume 
$\;v_N\;$ each, defined by the bag model. This renormalizes the volume
$\;V\rightarrow\xi V\;$ by the factor $\;\xi\equiv V_0/V =1 -\rho_Nv_N\;$,
where $\;\rho_N\equiv N_N/V\;$ is the nucleon density. Then, e.g., the baryon 
density takes the form
$$ n_B=\xi\int\left [
\frac{1}{3}\zeta_q\left ( 1 -\frac{2\alpha_s}{\pi}\right )
n_q(\stackrel{\rightarrow}{k}) + \zeta_Nn_N(\stackrel{\rightarrow}{k})
\right ]\frac{d\stackrel{\rightarrow}{k}}{(2\pi)^3} , $$
in which, in addition to the geometrical interaction taken into consideration 
by the factor $\;\xi\;$, the perturbative quark--quark interaction is 
simulated by the factor $\;1-2\alpha_s/\pi\;$. The Fermi distributions
$\;n_i(\stackrel{\rightarrow}{k})\;$ contain the free spectra 
$\;\omega_i(\stackrel{\rightarrow}{k})=\sqrt{k^2+m_i^2}-\mu_i\;$. The energy
density reads
$$ \varepsilon =\varepsilon_q + \varepsilon_N + B , $$
$$ \varepsilon_q =\xi\zeta_q\left ( 1 -\frac{2\alpha_s}{\pi}\right )
\int\varepsilon_q(\stackrel{\rightarrow}{k})n_q(\stackrel{\rightarrow}{k})
\frac{d\stackrel{\rightarrow}{k}}{(2\pi)^3} , $$
$$ \varepsilon_N =\xi\zeta_N
\int\varepsilon_N(\stackrel{\rightarrow}{k})n_N(\stackrel{\rightarrow}{k})
\frac{d\stackrel{\rightarrow}{k} }{(2\pi)^3} , $$
where the nonperturbative energy is given by a bag constant $\;B\;$.

Comparing the grand potentials for this model with $\;B^{1/4} =171\;MeV\;$ 
and those of pure quark and of pure nucleon phases, it was found that the
mixed clustered matter is more profitable at $\;n_B>8n_{0B}\;$ and
$\Theta \leq 50\;MeV\;$. Of course, one should not take too seriously the
numerical predictions of this model which is too oversimplified to be
realistic. For instance, pions and gluons are not considered here although 
they should play an important role in the deconfinement transition. Thus, 
pion fields make a nucleon bag unstable at quite moderate density and
temperature [74,75] close to those discussed in Introduction.

The most valuable in the clustered quark model is the idea of nonuniform
vacuum containing such local fluctuations that nucleons can be formed 
floating in the surrounding quark matter.

\section{String--Flip Potential Model}

An extreme realization of the dynamical confinement scenario is given by the 
string--flip potential model [76-81] based on the quantum--mechanical 
Hamiltonian
$$ H(\sr_1,\ldots ,\sr_N)=\sum_{i=1}^{N}\left ( -\frac{\nabla_i^2}{2m}\right )
+ V(\sr_1,\ldots ,\sr_N) $$
of nonrelativistic quarks. In the interaction term
$$ V(\sr_1,\ldots ,\sr_N) =\min\sum_{<ij>}\Phi(r_{ij}) \qquad 
(r_{ij} \equiv |\sr_i -\sr_j|) $$
the minimum is taken over all ways to group $\;N\;$ quarks into hadrons, so 
that the summation actually goes only over $\;<i,j>\;$ pertaining to the same 
hadron consisting of $\;2,3,6,9,12\;$ or $\;15\;$ quarks. Using the 
variational wave function
$$ \Psi (\sr_1,\ldots,\sr_N) =
\exp\{-\bt V(\sr_1,\ldots,\sr_N)\}D(\sr_1,\ldots,\sr_N) , $$
in which $\;\bt\;$ is a variational parameter and $\;D(\ldots)\;$ is a free 
Fermi gas Slater determinant, one minimizes the energy 
$\;E(\bt) =(\Psi,H\Psi)/(\Psi,\Psi)\;$ with respect to $\;\bt\;$. This gives
$\;\bt=\bt(\rho)\;$ as a function of density.

Variational Monte Carlo calculations for this model have been accomplished 
with the harmonic, $\;\Phi(r)=\frac{1}{2}m\om^2r^2\;$, and linear, 
$\;\Phi(r)=\sgm r\;$, confining potentials. The wave function has been 
symmetrized with respect to all of the quark coordinates, including these 
pertaining to different hadrons. This takes into consideration the exchange 
quark energy which is found [82] to be about $\;100\;MeV\;$.

The results show that the quarks always coalesce into the lowest energy set 
of flux tubes, which is characteristic of an adiabatic approximation to the 
strong coupling limit of $\;QCD\;$. At low densities $\;\rho<\rho_0\;$ quarks 
cluster into isolated hadrons. As the density increases, the value of 
$\;\bt(\rho)\;$ decreases, first slowly, but at $\;\rho\approx 1.5\rho_0\;$ 
exponentially. For $\;\rho\geq 2\rho_0\;$ the wave functions of separate 
hadrons strongly overlap, which may be interpreted as a transition to the 
quark matter, although formally all quarks are yet confined. The complete 
deconfinement occurs at asymptotically high densities, when 
$\;\bt(\rho)\ra 0\;$, and the wave function becomes that
of a free Fermi gas of quarks.

The quark--quark pair correlation function was calculated [76,79] using 
Monte Carlo techniques and the longitudinal response function, using 
molecular dynamics simulations [83]. At low densities these functions have 
the properties typical of a set of isolated hadrons. For instance, the 
pair correlation function has a sharp peak displaying strong correlations 
between quarks. The form of this function is similar to that of liquids 
[84,85]. As the density increases, the peak in the pair correlation function 
relaxes smoothly to that of a Fermi gas of quarks, disappearing completely 
as $\;\rho\ra\infty\;$.

Thus, the string--flip model demonstrates a smooth transition from hadron 
matter at low densities to a free Fermi gas of quarks at high densities. 
Quarks at all densities are to some extent confined becoming an 
absolutely free gas only in the limit $\;\rho\ra\infty\;$.

The main difficulty in dealing with this model is the necessity of the 
minimization of the Hamiltonian interaction term over all strings connecting 
quarks, which involves complicated Monte Carlo calculations. This procedure 
is necessary for eliminating colour van der Waals forces between colour 
singlet hadrons. These forces  are present in confining two--body 
potentials and, if not eliminated, produce large spurious energies in 
nuclear matter.

However, in a consistent statical approach this kind of divergence can 
be automatically cancelled if one takes into account the corresponding 
correlation functions, called smoothing or screening functions, which 
smooth the interaction potentials in the regions of their divergence 
[86-88]. It is possible, starting with divergent interaction potentials, 
to construct regular iterative theory [35,89] for Green functions, whose 
each step involves only smoothed potentials containing no divergencies. 
In the lower orders of this correlated interaction theory [35] the 
situation is such that could be obtained by replacing from the beginning 
the bare interaction potential by a smoothed one, that is by defining an 
effective Hamiltonian with a pseudopotential instead of the initial 
potential [86]. Such a replacement is an approximation neglecting some double
and all triple correlations [35].

The Hamiltonian of the string--flip model including the minimization over 
strings is an effective Hamiltonian, in which this minimization plays the 
role of smoothing. In lieu of this complicated minimization requiring 
Monte Carlo techniques, it is equival ent to replace the confining 
potential by a smoothed one.

R\"opke at al. [90-93] have developed a many--body approach to quark--nuclear 
matter generalizing and simplifying the string--flip model. This approach
starts with an effective Hamiltonian
$$ H_{eff}(\sr_1,\ldots,\sr_N)=\sum_{i=1}^N\left ( m_i  -
\frac{\nabla_i^2}{2m_i}\right ) +\frac{1}{2}\sum_{i,j}^N V_{eff}(\sr_{ij}) , $$
in which $\;m_i\;$ stands for the quark masses, and the effective interaction
$\;V_{eff}(\sr)\equiv s(r)\Phi(r)\;$ is screened by a function $\;s(r)\;$
defining the probability that two quarks at distance $\;r\;$ are next 
neighbours. The screening function is to be found from some additional 
equations. In the case of independent particles with density $\;\rho\;$, 
one has $\;s(r)=\exp ( -\frac{4\pi}{3}\rho r^3)\;$. Inside a hadron, when 
$\;r\ll a\;$, where $\;a\equiv\rho^{-1/3}\;$ is an average interquark 
distance in the system, we have $\;s(r)\simeq 1\;$, so quarks interact 
through a bare confining potential. For the quarks pertaining to different 
hadrons, i.e. when $\;r\gg a\;$, we get $\;s(r)\simeq 0\;$, which is called 
the saturation property [91].

The many--body approach to clustering quark matter [90-93] makes it possible 
to give a unique description of both the clustered hadrons as well as the 
free--quark phase in the same way as it has been done for the nuclear matter 
whose nucleons could form the deuteron, triton , and $\;\al\;$--particle bound
states [94,95]. The total density of clustering matter is a sum $\;\rho=\sum_i
z_i\rho_i\;$, in which $\;z_i\;$ is a compositeness factor and $\;\rho\;$ is 
the density of an $\;i\;$--component of clusters composed of $\;z_i\;$ 
particles each.

In the many--body approach the quark system may contain both quasifree quarks
(scattering states) and various clusters (bound states). At zero temperature,
nucleons start dissolving at $\;n_B\approx 4.3n_{0B}\;$. At zero baryon
density the transition of hadron matter to free--quark phase starts at 
$\;\Theta_d\approx 200\;MeV\;$ where some of meson states are already 
dissolved. But other bound quark states do exist yet at this temperature. 
The dissolution of the bound states occurs gradually with temperature. 
Some mesons survive up to $\;\Theta\approx 2\Theta_d\;$. The gradual 
dissociation of bound states above $\;\Theta_d\;$ is in agreement with 
the dynamical deconfinement scenario. 

Although the many--body approach provides the principal possibility of 
treating bound states together with quasifree ones, it has the following 
technical  obstacles. Each bound state is described by a 
Bethe--Goldstone--type equation for a two--or three--particle Green 
function, depending on the compositeness of this bound state. This type 
of equations, as is known, is very difficult to deal with. When there are 
many bound states interacting with each other as well as
with unbound particles, then one has to deal with a system of 
many interrelated Bethe--Goldstone equations, in addition to several 
equations for one--particle Green functions. In such a case the problem 
becomes practically as difficult as chromodynamics itself.

\section{Concept of Cluster Coexistence}

As follows from the previous Sections, there is need of such a statistical 
model that could provide a realistic description of clustering matter being 
at the same time treatable. One almost evident simplification would be, 
instead of suffering with a system of Bethe--Goldstone equations, to 
consider bound clusters as separate objects. This can be understood as a 
kind of renormalization by integrating out the internal degrees of freedom 
corresponding to the motion of particles inside each of clusters, so that 
only the center--of--mass degrees of freedom are left.

Enumerate all possible cluster states, including those of unbound, or free,
particles, by the index $\;i\;$. Let $\;\psi_i(\sr)=[\psi_i^\al(\sr)]\;$ be 
the field operator of an $\;i\;$--cluster; this operator being a column 
in the space of quantum degrees of freedom indexed by  $\;\al\;$. The 
quantum degrees of freedom include spin, isospin, colour and flavour 
indices. The field operators satisfy the commutation relations.
$$ [\psi_i^\al(\sr),\psi_j^\ga(\sr')]_{\mp} = 0 , \qquad
[\psi_i^\al(\sr),\stackrel{\dagger}{\psi_j^\ga}(\sr')]_{\mp} =
\delta_{ij}\delta_{\al\ga}\delta(\sr-\sr') , $$
where the upper sign stands for Bose and the lower, for Fermi statistics.

Take an effective Hamiltonian in the form
\be
H = \hat E -\sum_{i}\mu_i\hat N_i ,
\ee
in which the energy operator
\be
\hat E =\sum_i\hat E_i + \sum_{i,j}\hat E_{ij}
\ee
consists of the single--cluster energies
\be
\hat E_i = \int {\psi}_i^\dgr(\sr)K_i(\stackrel{\ra}{\nabla})
\psi_i(\sr)d\sr ,
\ee
where $\;K_i(\stackrel{\ra}{\nabla})\;$ is a kinetic--energy operator, and 
of the interaction energies
\be
\hat E_{ij} =\frac{1}{2}\int\psi_i^\dgr(\sr)\psi_j^\dgr(\sr')\Phi_{ij}(\sr -
\sr')\psi_j(\sr')\psi_i(\sr)d\sr d\sr' ,
\ee
with the interaction potentials $\;\Phi_{ij}(\sr)\;$ having the symmetry 
property
\be
\Phi_{ij}(\sr) =\Phi_{ij}(-\sr) = \Phi_{ji}(\sr) .
\ee
In the second term of (1), $\;\mu_i\;$ is the chemical potential of 
$\;i\;$-clusters and
\be
\hat N_i =\int \psi_i^\dgr(\sr)\psi_i(\sr)d\sr
\ee
is a number--of--cluster operator.

The grand potential 
\be
\Om = -\Theta\ln {\rm Tr}\; e^{-\bt H}
\ee
defines the pressure
\be
p \equiv -\frac{\Om}{V} =\frac{\Theta}{V}\ln {\rm Tr} \; e^{-\bt H} .
\ee
The energy density is given by
\be
\ep \equiv \frac{1}{V}\lgl \hat E\rgl ,
\ee
where $\;\lgl\ldots\rgl\;$ denoted the statistical averaging.

For each kind of clusters we may define their density
\be
\rho_i(\sr) =\lgl\psi_i^\dgr(\sr)\psi_i(\sr)\rgl ,
\ee
the number of clusters
\be
N_i\equiv \lgl\hat N_i\rgl =\int\rho_i(\sr)d\sr ,
\ee
and the average density
\be
\rho_i\equiv \frac{N_i}{V} =\frac{1}{V}\int\rho_i(\sr)d\sr .
\ee

The number of constituents of an $\;i\;$--cluster is called the compositeness 
factor $\;z_i\;$. The total number of elementary particles is
\be
N = \sum_i z_iN_i ,
\ee
so that the average density of the system can be written as
\be
\rho\equiv \frac{N}{V} = \sum_i z_i\rho_i .
\ee

The chemical potentials $\;\mu_i\;$ can be defined from the conservation 
laws accepted for the given system. For example, if the numbers $\;N_i\;$ 
for each sort of clusters are fixed, then (12) gives 
$\;\mu_i(\Theta,\rho_i)\;$. If the total number of constituents (13) 
is fixed, then from the equilibrium condition $\;\delta\Om =0\;$ we have
\be
\frac{\mu_i}{z_i} = \frac{\mu_j}{z_j} ,
\ee
which, together with (14), defines $\;\mu_i(\Theta,\rho)\;$. When neither 
$\;N_i\;$ nor $\;N\;$ are fixed, then $\;\mu_i=0\;$.

An important quantity is the {\it cluster probability}
\be
w_i \equiv z_i\frac{\rho_i}{\rho} = z_i\frac{N_i}{N}
\ee
enjoying the conditions
\be
0 \leq w_i \leq 1 , \qquad \sum_i w_i = 1 ,
\ee
following from (14). With (12), we may rewrite (16) as
\be
w_i =\frac{z_i}{N}\int\rho_i(\sr)d\sr .
\ee

If the system of particles can form a number of bound states, then how could 
we define numerous interactions between clusters? The number of such 
interactions can be drastically reduced for those clusters whose interaction 
potentials $\;\Phi_{ij}(\sr)\;$ have the same type of behaviour at large 
distance, for example, diminishing as $\;r\;$ increases. Consider such 
clusters with a similar behaviour of $\;\Phi_{ij}(\sr)\;$. Let two clusters, 
$\;m\;$ and $\;n\;$, coalesce into one, $\;i\;$, so that the compositeness 
numbers involved in the reaction $\;m+n\ra i\;$ satisfy the obvious relation 
$\;z_m+z_n=z_i\;$. And let aside be another cluster--spectator, $\;j\;$, 
as is shown in Fig.1. Assume that the coalescence of the cluster--actors 
does not influence the cluster--spectator, in the sense that its interaction 
with the initial two clusters is the same as with that one formed after the 
coalescence:
\be
\Phi_{mj}(\sr) +\Phi_{nj}(\sr) = \Phi_{ij}(\sr) .
\ee
From this assumption, using the conservation law $\;z_m+z_n = z_i\;$, we find
the relation
\be
\frac{\Phi_{ij}(\sr)}{z_iz_j} = \frac{\Phi_{mn}(\sr)}{z_mz_n} ,
\ee
permitting to express the interaction potentials of different clusters 
through one calibration potential.

\section{Screening of Interaction Potentials}

The interaction potential $\;\Phi_{ij}(\sr)\;$ may be divergent for some 
regions of $\;\sr\;$. However, it would not be correct to replace this 
potential by a smoothed, or screened, one just in the Hamiltonian (1). 
Such a screening appears in the process of decoupling of binary propagators 
with taking account of interparticle correlations, as is done in the 
correlated iteration theory [35]. If one wishes to write an effective 
Hamiltonian corresponding to a given correlated approximation, it is not 
sufficient to merely rearrange the operator terms of the Hamiltonian, but a 
nonoperator correcting term must be added.

Let us illustrate this for the correlated mean--field approximation [35] 
when the Hamiltonian (1) may be presented in the form
$$ H = \sum_i H_i +CV , $$
\be
H_i =\int \psi_i^\dgr(\sr)\left [ K(\stackrel{\ra}{\nabla}) + U_i(\sr) -
\mu_i \right ]\psi_i(\sr)d\sr ,
\ee
in which the correlated mean field
\be
U_i(\sr) =\sum_j\int \stackrel{-}{\Phi}_{ij}(\sr -\sr')\rho_j(\sr')d\sr'
\ee
contains the screened potential
\be
\stackrel{-}{\Phi}_{ij}(\sr) =s_{ij}(\sr)\Phi_{ij}(\sr)
\ee
smoothed by the smoothing function having the symmetry property
\be
s_{ij}(\sr) =s_{ji}(\sr) = s_{ij}(-\sr) .
\ee
The nonoperator correcting term in (21) is $\;CV\;$, and this cannot be 
put zero, as will be shown below.

Since the exact Hamiltonian (1) does not depend on cluster densities, 
varying the grand potential (7) with respect to $\;\rho_i(\sr)\;$ we have
\be
\frac{\delta\Om}{\delta\rho_i(\sr)} = 
\left \lgl \frac{\delta H}{\delta\rho_i(\sr)}\right \rgl = 0 .
\ee
In order that the exact Hamiltonian (1) would be correctly represented by 
the approximate Hamiltonian (21) requires that the latter must satisfy (25), 
which
yields 
\be
\frac{\delta C}{\delta\rho_i(\sr)} + \frac{1}{V}\sum_j\left \lgl
\frac{\delta H_j}{\delta\rho_i(\sr)}\right \rgl = 0 .
\ee
Substituting here
$$ \left \lgl \frac{\delta H_i}{\delta\rho_i(\sr)}\right \rgl = 
\int\frac{\delta U_j(\sr')}{\delta\rho_i(\sr)}\rho_i(\sr')d\sr' , $$
transforms (26) into
\be
\frac{\delta C}{\delta\rho_i(\sr)} + \frac{1}{V}\sum_j\int
\frac{\delta U_j(\sr')}{\delta\rho_i(\sr)}\rho_j(\sr')d\sr' = 0 .
\ee

If the smoothing function $\;s_{ij}(\sr)\;$ does not depend on 
$\;\rho_i(\sr)\;$, then (22) gives
$$ \frac{\delta U_i(\sr)}{\delta \rho_j(\sr')} = \bar \Phi_{ij}(\sr -\sr') . $$
And Eq.(27) becomes
$$ \frac{\delta C}{\delta\rho_i(\sr)} + \frac{1}{V}
\sum_j\int\stackrel{-}{\Phi}_{ij}(\sr -\sr')\rho_j(\sr')d\sr' = 0 . $$
The solution of the latter variational equation, up to a constant that can 
be omitted, is
$$ C = -\frac{1}{2V}\sum_{i,j}\int
\stackrel{-}{\Phi}_{ij}(\sr -\sr')\rho_i(\sr)\rho_j(\sr')d\sr d\sr' . $$
Note that the Hamiltonian (21) with the obtained correcting term could be 
derived from (1) with the substitution
$$ \psi_i^\dgr(\sr)\psi_j^\dgr(\sr')\psi_j(\sr')\psi_i(\sr) \ra $$
$$ \ra
s_{ij}(\sr -\sr')\left \{ \psi_i^\dgr(\sr)\psi_i(\sr)\rho_j(\sr') +
\rho_i(\sr)\psi_j^\dgr(\sr')\psi_j(\sr') - \rho_i(\sr)\rho_j(\sr')\right\}, $$
corresponding to the correlated Hartree approximation [35].

In general, the smoothing function $\;s_{ij}(\sr)\;$ is dependent on 
$\;\rho_i(\sr)\;$. Therefore, the correcting term is to be found from Eq.(26) 
or (27). In any case, the correcting term depends on thermodynamic variables 
through the densities $\;\rho_i(\sr)\;$. Thus, neglecting this term would 
disfigure the correct statistical description, and the behaviour of 
thermodynamic functions could be completely spoiled.

Let us pass to a uniform system when $\;\rho_i(\sr)=\rho_i\;$. In principle, 
a multicomponent system can display a variety of nonuniform states related to 
the solidification of one or several components. For example, an ensemble of 
fully ionized nuclei can form a crystalline lattice immersed in a uniform 
electron background, which models the high--density matter of white 
dwarfs [96]. It may be that some heavy cluster components crystallize 
while others are liquid--like as it happens in superionic conductors [97]. 
It also may be that heavy clusters form an amorphous solid while light ones 
move in a conduction band as in glassy metals [98]. In the cores of neutron 
stars a Gibbs mixture can exist when in some volumes of space a lattice 
structure appears while others are filled by a liquid--like nuclear 
matter [99]. We leave aside all these possibilities considering in what 
follows only uniform systems.

In the uniform case, the mean field (22) becomes
\be
U_i(\sr) =\sum_j\Phi_{ij}\rho_i \equiv U_i
\ee
with the interaction integral
\be
\Phi_{ij} \equiv \int \stackrel{-}{\Phi}_{ij}(\sr)d\sr .
\ee
Instead of (25), we have
\be
\left\lgl\frac{\delta H}{\delta \rho_i}\right\rgl = 0 .
\ee
The correcting equation (27), defining the correcting term, changes to
\be
\frac{\delta C}{\delta \rho_i} + 
\sum_j\frac{\delta U_j}{\delta\rho_i}\rho_j = 0 .
\ee

The field operators, for a uniform system, can be expanded in plane waves,
$$ \psi_i(\sr) = \frac{1}{\sqrt{V}}\sum_k a_i(\sk)e^{i\sk\sr} ; \qquad
a_i(\sk) = \left [ a_i^\al(\sk)\right ] . $$
Then an $\;i\;$--cluster Hamiltonian in (21) transforms to 
\be
H_i =\sum_k\om_i(\sk)a_i^\dgr(\sk)a_i(\sk) ,
\ee
with the effective spectrum
\be
\om_i(\sk) \equiv \ep_i(\sk) -\mu_i , \qquad \ep_i(\sk) \equiv K_i(\sk) +U_i .
\ee
The number of clusters (11) reads
\be
N_i =\sum_k\lgl a_i^\dgr(\sk)a_i(\sk)\rgl =\zeta_i\sum_k n_i(\sk) ,
\ee
where $\;\zeta_i\;$ is a degeneracy factor, i.e. the number of quantum 
states, and
\be
n_i(\sk) =\left\{ \exp\left [ \bt\om_i(\sk)\right ]\mp 1\right \}^{-1}
\ee
is the Bose (upper sign) or Fermi (lower sign) momentum distribution. The 
cluster density (12) is
\be
\rho_i =\zeta_i\int n_i(\sk)\frac{d\sk}{(2\pi)^3} .
\ee

For the grand potential (7) one gets
$$ \Om =\sum_i\Om_i +CV , $$
\be
\Om_i =\mp \Theta V\zeta_i\int \ln\left [ 1 \pm n_i(\sk)\right ] 
\frac{d\sk}{(2\pi)^3} .
\ee
The pressure (8) writes
$$ p = \sum_i p_i - C , $$
\be
p_i = \pm\Theta\zeta_i\int\ln\left [ 1 \pm n_i(\sk)\right ]
\frac{d\sk}{(2\pi)^3} .
\ee
The energy density (9) is
$$ \ep =\sum_i\ep_i + C , $$
\be
\ep_i =\zeta \int\ep_i(\sk)n_i(\sk)
\frac{d\sk}{(2\pi)^3} .
\ee
The cluster probability (18) becomes
\be
w_i =\frac{z_i}{\rho}\zeta_i\int n_i(\sk)
\frac{d\sk}{(2\pi)^3} .
\ee

An additional simplification comes for isotropic systems, which is usually 
assumed, when $\;\ep_i(\sk)=\ep_i(k)\;$, where $\;k\equiv|\sk|\;$. Then 
$\;\om_i(\sk)=\om_i(k)\;$ and $\;n_i(\sk) = n_i(k)\;$. If the spectrum
$\;\om_i(k)\;$ is such that the asymptotic properties
$$ k^3\ln\om_i(k)\ra 0 \qquad (k\ra 0) , $$
$$ \om_i(k)\ra\infty \qquad (k\ra\infty ) $$
hold true, then the cluster pressure and the cluster energy density reduce to
$$ p_i =\frac{\zeta_i}{6\pi^2}\int_0^\infty k^3\ep_i'(k)n_i(k)dk , $$
$$ \ep_i =\frac{\zeta_i}{2\pi^2}\int_0^\infty k^2\ep_i(k)n_i(k)dk , $$
where $\;\ep_i'(k)\equiv d\ep_i(k)/dk\;$.

These are the basic formulas which we shall use in what follows.

\section{Coexisting Multiquark Clusters}

The concept of cluster coexistence has been applied first to nuclear matter 
consisting of different multiquark clusters. The interest to this problem was
motivated by the A.M.Baldin commulative effect [1] and the related discussion 
of the possible existence of multiquark clusters in nuclei [100,101].

Since at high density or temperature relativistic effects play an important 
role, the kinetic term of the cluster spectra is taken in the relativistic 
form $\;K_i(\sk)=\sqrt{k^2+m_i^2}\;$, where $\;m_i\;$ is the cluster mass. A 
system of $\;3\;$--, $\;6\;$--, $\;9\;$--, and $\;12\;$--quark clusters has 
been considered in the excluded--volume approximation [102-106]. The 
$\;3\;$-- and $\;9\;$--quarks are Fermions, and the $\;6\;$-- and 
$\;12\;$--quarks are Bosons. The Bosons with the lowest mass, that is the 
$\;6\;$--quarks, can drop down in the Bose--Einstein condensate, when 
$\;\om_6(0)=0\;$, which fixes $\;\mu_6\;$. The chemical potentials always 
satisfy (15).

In the excluded--volume approximation the interaction between clusters is 
considered geometrically by putting $\;\Phi_{ij}\;$ zero, but replacing the 
total volume $\;V\;$ by the free volume $\;V_0\;$,
\be
V \ra V_0 \equiv V -\sum_iN_iv_i ,
\ee
where $\;v_i\;$ are cluster volumes. This reduces the volume
\be
V \ra \xi V ; \qquad \xi \equiv \frac{V_0}{V} = 1 -\sum_i\rho_i v_i ,
\ee
by a factor $\;\xi\in [0,1]\;$. Equivalently, this can be interpreted as a 
reduction of the degeneracy factor
\be
\zeta_i \ra \tilde\zeta_i \equiv \xi\zeta_i .
\ee
The reduction factor in (42) and (43) can also be written as 
$$ \xi =\left ( 1 + \sum_i\rho_i^{(0)}v_i\right )^{-1} ; \qquad
\rho_i^{(0)} \equiv \zeta_i\int n_i(k) \frac{d\sk}{(2\pi)^3} . $$

Thus, for the density of clusters, pressure, and energy density one has
$$ \rho_i =\tilde\zeta_i\int n_i(k)\frac{d\sk}{(2\pi)^3} , $$
$$ p =\pm\Theta\sum_i\tilde\zeta_i\int\ln\left [ 1 \pm n_i(k)\right ]
\frac{d\sk}{(2\pi)^3} , $$
\be
\ep =\sum_i\tilde\zeta_i\int\ep_i(k)n_i(k)\frac{d\sk}{(2\pi)^3} .
\ee

The volumes of clusters are supposed to be related by the equation
\be
\frac{v_i}{m_i} = \frac{v_j}{m_j} .
\ee
This permits to express all cluster volumes $\;v_i =m_iv_3/m_3\;$ through the
$\;3\;$--quark volume $\;v_3=4\pi r_3^3/3\;$ with the radius $\;r_3=0.4\;fm\;$
of the nucleon core.

The used multiquark parameters are given in Table 1. The $\;3\;$--quark is a 
nucleon. The $\;6\;$--quark mass $\;m_6=1944\;MeV\;$ corresponds to an average
value over the masses of several light narrow dibaryons that are claimed to 
be observed in experiments [107]. The mass $\;m_6=2163\;MeV\;$ is taken from 
the bag--model calculation of Jaffe [108] and the $\;9\;$-- and 
$\;12\;$--quark parameters are elicited from the bag model of Matveev and 
Sorba [109].

The $\;6\;$--quark probability depends on the value of the mass $\;m_6\;$ as 
is illustrated in Fig.2. The results for other cluster probabilities are 
displayed in Figs.3 and 4, where $\;m_6=2163\;MeV\;$. The probabilities of 
heavy clusters are always very small: $\;w_9 < 0.1\;$, $\;w_{12} < 0.01\;$.

As far as we limited here by the length of this review, we shall not discuss 
in detail the results of our calculations, which can be found in the cited 
papers. We think that the presented figures speak for themselves: it is 
better to see once than to listen hundred times.

The coexistence of nucleons with $\;6\;$--quark clusters has been considered 
as well in the mean--field approximation [110] with the effective interaction 
potential
$$ \Phi_{ij}(\sr) =2\pi\frac{a_{ij}}{m_{ij}}\delta(\sr) -
\frac{\al_\pi}{r}\exp\left ( -m_\pi r\right ) . $$
The first term here is the Fermi pseudopotential for the core interaction 
with the scattering length $\;a_{ij}\equiv\frac{1}{2}(a_i+a_j)\;$, where 
$\;a_i\equiv a_{ii}\;$, and the reduced mass 
$\;m_{ij}\equiv m_im_j/(m_i+m_j)\;$. The second term is caused 
by the one--pion exchange; $\;\al_\pi=0.08\;$ being the pion coupling 
parameter and $\;m_\pi=140\;MeV\;$, the pion mass. Again, to reduce 
the number of model parameters, the relation $\;a_i^3/m_i=a_j^3/m_j\;$, 
similar to (45), is accepted. Then, all scattering lengths
$\;a_i=a_3(m_i/m_3)^{1/3}\;$ are expressed through the nucleon scattering 
length $\;a_3=1.6fm\;$. The $\;6q\;$--probability is shown in Fig.5.

Note that the mean--field approximation is valid if $\;|U_i|\ll m_i\;$, 
which is true for the densities up to about $\;10\rho_0\;$.

The models of this Section serve rather as a qualitative illustration of 
cluster coexistence. They can have sense only at temperatures and densities 
much lower than those characteristic of deconfinement, as unbound quarks are 
not included here.

\section{Baryon Rich Matter} 

Include into consideration unbound (quasifree) quarks that can, in principle, 
coexist with multiquark clusters. Whether and when quasifree quarks really 
coexist should be determined in a self--consistent way from the conditions of 
thermodynamic advantageousness and stability. To compare the results with 
those of the previous section, consider again the case of baryon rich matter, 
when $\;\rho=3n_B\;$, that is when the generation of particles from vacuum 
can be neglected. 

Denote the chemical potential of a quark by $\;\mu\equiv\mu_q\;$. Then 
relation (15) yields $\;\mu_i=z_i\mu\;$.

The strengths of characteristic interactions between baryons and between 
quarks are of the order of or higher than the expected deconfinement 
temperature, so these interactions must be taken into account. The mean 
field acting on quarks may be written in a bag--model--motivated form [32] 
as $\;U_q=BV/N=B/\rho\;$. This gives for quarks the spectrum
\be
\ep_q(k)=\sqrt{k^2+m_q^2}+\frac{B}{\rho} .
\ee
The mean--field term in (46) contains the total quark density $\;\rho\;$, 
which means that each quasifree quark interacts in the same way with other 
unbound quarks as well as with quarks entering into bound clusters. 
The interaction potentials between different baryons can be expressed, basing 
on relation (20), through the nucleon--nucleon interaction potential 
$\;\Phi_{33}(\sr)\;$,
\be
\Phi_{ij}(\sr) =\frac{z_iz_j}{9}\Phi_{33}(\sr) .
\ee
There are several such effective potentials obtained from nucleon--nucleon 
scattering experiments [111] or from analysing the deuteron properties [112]. 
We opt for the Bonn potential [113]. The common consensus is that 
thermodynamics of nuclear matter does not depend on the mutual orientation 
of spins of interacting nucleons. Averaging over spin directions nullifies 
the spin terms of the interaction potential. The so--called cut--off terms 
of the Bonn potential can be neglected, since they start playing an essential 
role only for very short distances $\;\leq 0.1fm\;$, which would correspond 
to the baryon density $\;n_B\geq 10^3n_{0B}\;$. We assume that the interaction
between any pair of nucleons is the same, because of which in the isospin 
term of the Bonn potential we put the total isospin $\;I_1+I_2 =1\;$ 
describing the interaction between protons or neutrons. The so obtained 
radial part of the Bonn potential [113] reads
\be
\Phi_{33}(\sr) =\sum_{i=1}^4\frac{\al_i}{r}\exp\left ( -\ga_ir\right )
\ee
with the parameters
$$ \al_1=16.7, \quad \al_2=2.7, \quad \al_3=-7.8, \quad \al_4 =-2.7 , $$
$$ \ga_1=738\;MeV, \quad \ga_2=769\;MeV, \quad \ga_3 =550\;MeV, \quad 
\ga_4 =983\;MeV.$$
The interaction potential (48) is integrable, thus, it does not necessarily
require the smoothing procedure [35,88,114]. For the interaction--energy 
density (29) we have
\be
\Phi_{33} =\int\Phi_{33}(\sr)d\sr ,
\ee
which yields $\;\Phi_{33}=4.1\times 10^{-5}\;MeV^{-2}=315\;MeV\;fm^3\;$. 
Note that $\;\Phi_{33}\rho_0=164\;MeV\;$, hence 
$\;\Phi_{33}\rho_0 \approx E_0 =\rho_0^{1/3}\;$, from where 
$\;\Phi_{33}\approx \rho_0^{-2/3}\;$.

For the spectra of baryons we take
\be
\ep_i(k)=\sqrt{k^2+m_i^2} +\frac{z_i}{9}\left (\rho - \rho_q\right )\Phi_{33},
\ee
where $\;\rho_q\;$ is the quark density and $\;i\;$ enumerates multiquark 
clusters: $\;3q,\; 6q,\;9q\;$ $12q\;$, and so on. The masses of bound clusters
up to $\;z_i=12\;$ are taken from Table 1, with the six--quark mass 
$\;m_6=1944\;MeV\;$. For $\;z_i\geq 15\;$ we use the formula 
$\;m_i\approx (z_i/3)m_3\;$ for the masses of heavy multiquark clusters [22].

For quarks we accept the mass $\;m_q=7\;MeV\;$ and the degeneracy factor 
$\;\zeta_q=12\;$ corresponding to spin $1/2,\; N_c=3\;$, and $\;N_f=2\;$. 
The bag constant $\;B^{1/4}=235\;MeV\;$.

The results of numerical calculations [115-119] are presented in Figs.6-13. 
At $\;\Theta=0\;$ and $\;\rho=\rho_0\;$, the $\;6q\;$--probability is 
$\;w_6=0.18\;$, which agrees with the estimates of the $\;6q\;$--admixture in 
nuclei [120]. The  heavy--multiquark probabilities are always small: 
$\;w_9 <10^{-3}, \; w_{12} < 10^{-5}\;$, and $\;w_{15} < 10^{-7}\;$. At zero 
temperature, only the  Bose--condensed $\;6q\;$--clusters exist, the 
probabilities of heavier ones being strictly zero. Unbound quarks, at 
$\;\Theta=0\;$, are absent below the density $\;\rho_q^{nuc}\approx 2\rho_0\;$
when they start appearing. This is why the characteristic density 
$\;\rho_q^{nuc}\;$ may be called the {\it nucleation density}. As we see, 
the value of the latter agrees with the corresponding estimates from 
Introduction.

The stability of the quark--baryon mixture is controlled by checking the 
minimum of the free energy $\;F =\Om +\sum_i\mu_iN_i\;$ whose density can 
be written as
$$ f \equiv\frac{F}{V} =\frac{\Om}{V} +\sum_i\mu_i\rho_i = -p +\mu\rho , $$
and also by requiring the validity of the stability conditions [54]
$$ -\Theta\frac{\partial^2f}{\partial\Theta^2} > 0 , \qquad 
\rho\frac{\partial p}{\partial\rho} > 0 . $$

The probability of unbound quarks increases, with temperature or density 
monotonically showing that the deconfinement is a gradual crossover but not 
a sharp transition. This is in agreement with numerical simulations [121] on 
$\;16^3\times 24\;$ lattice, for $\;N_f=2\;$, which has demonstrated that 
quark--quark correlation functions at $\;\Theta\approx 1.5\Theta_d\;$ are 
very similar to the zero--temperature wave functions of the corresponding 
particles.

\section{Zero Baryon Density} 

The case opposite to that of the previous section is when the baryon density 
is zero, $\;n_{0B}=0\;$, and all particles are generated from vacuum. Then 
$\;\mu_i=0\;$. In this case, we can compare our calculations with the lattice 
numerical simulations that are available only for $\;n_B=0\;$

The spectra of gluons and quarks are again taken in the form
\be
\om_g(k) = k +\frac{B}{\rho} , \qquad 
\om_q(k) =\sqrt{k^2+m_q^2}+\frac{B}{\rho}
\ee
with the bag--motivated mean fields. The interaction of hadrons is considered 
in the excluded--volume approximation. The cluster volumes 
$\;v_i=m_iv_2/m_2\;$, according to (45), are expressed through the volume 
$\;v_2\equiv 4\pi r^3_2/3\;$ of the lightest cluster with $\;z_i=2\;$. 
The bag constants for the $\;SU(2)\;$ and $\;SU(3)\;$ systems are to be 
different [122] with the relation $\;B_{SU(2)}\approx 0.4B_{SU(3)}\;$.

The presentation of results is convenient to perform in relative quantities 
reduced to these of a reference system. The role of such a system is naturally
played by the Stefan--Boltzmann quark--gluon gas. The latter, by definition, is
an ensemble of free massless quarks, antiquarks, and gluons. The pressure and 
energy density of the Stefan--Boltzmann quark--gluon plasma are, respectively,
\be
p_{SB} = p_q^{(0)} + p_{\bar q}^{(0)} + p_g^{(0)} , \qquad
\ep_{SB} =\ep_q^{(0)} + \ep_{\bar q}^{(0)} +\ep_g^{(0)} ,
\ee
where
$$ p_i^{(0)} =\pm\Theta\zeta_i\int\ln\left [ 1 \pm n_i^{(0)}(k)\right ]
\frac{d\sk}{(2\pi)^3} , $$
\be
\ep_i^{(0)} =\zeta_i\int kn_i^{(0)}(k)\frac{d\sk}{(2\pi)^3} ,
\ee
the index $\;i=q,\bar q,g\;$ enumerates quarks, antiquarks, and gluons with 
the momentum distribution
\be
n_i^{(0)}(k) =\left \{ \exp\left [ \bt (k-\mu_i)\right ]\mp 1\right \}^{-1} ,
\ee
in which the upper sign is for Bosons (gluons) and the lower one, for Fermions
(quarks and antiquarks); the chemical potentials being
\be
\mu_q = -\mu_{\bar q}\equiv \mu, \qquad \mu_g=0 .
\ee
Eqs. (54) and (55) permit to write down (53) as 
\be
p_i^{(0)} =\frac{\zeta_i}{6\pi^2}\int_0^\infty
\frac{k^3dk}{\exp [\bt (k-\mu_i)]\mp 1}, \qquad \ep_i^{(0)} =3p_i^{(0)} .
\ee
An exact integration yields
$$ p_q^{(0)} + p_{\bar q}^{(0)} = \frac{\zeta_q}{12}
\left ( \frac{7\pi^2}{30}\Theta^4 +\mu^2\Theta^2 +
\frac{\mu^4}{2\pi^2}\right ) , $$
\be
p_g^{(0)} =\frac{\pi^2}{90}\zeta_g\Theta^4 .
\ee
Thus, the Stefan--Boltzmann pressure is
\be
p_{SB} =\frac{\pi^2}{90}\left ( \zeta_g + \frac{7}{4}\zeta_q\right ) 
\Theta^4 + \frac{\zeta_q}{12}\mu^2\Theta^2
\left ( 1 +\frac{\mu^2}{2\pi^2\Theta^2}\right ) .
\ee
This is to be compared with the $\;QCD\;$ pressure corresponding to the 
$\;QCD\;$ grand potential, given in Section 3, for zero coupling $\;g=0\;$,
\be
p_{QCD} =\frac{\pi^2}{45}
\left ( N_c^2 - 1 +\frac{7}{4}N_fN_c\right )\Theta^4 + 
\frac{N_fN_c}{6}\mu^2\Theta^2\left ( 1 +\frac{\mu^2}{2\pi^2\Theta^2}\right ) .
\ee
Due to the relations for the degeneracy factor of gluons, 
$\;\zeta_g = 2\times (N_c^2-1)\;$, and for that of quarks, 
$\;\zeta_q=2\times N_f\times N_c\;$, and antiquarks, 
$\;\zeta_q=\zeta_{\bar q}\;$, we see that (58) and (59) coincide with each 
other. Therefore, the Stefan--Boltzmann plasma is the asymptotic 
high--temperature limit of quantum chromodynamics.

The baryon density for the Stefan--Boltzmann plasma is
\be
n_B\equiv \frac{1}{3}\left ( \rho_q - \rho_{\bar q}\right ) = 
\frac{\zeta_q}{3}\int \left [ n_q(k) - n_{\bar q}(k)\right ]
\frac{d\sk}{(2\pi)^3} .
\ee
Either calculating (60) directly or using the derivative $\;n_B =\partial p/
\partial\mu_B\;$, with $\;\mu_B=3\mu\;$ we have
\be
n_B =\zeta_q\frac{\mu}{18\pi^2}\left ( \mu^2 +\pi^2\Theta^2\right ) .
\ee
From here, one gets the equation
\be
\mu^3 +\pi^2\Theta^2\mu - \frac{18\pi^2}{\zeta_q}n_B = 0
\ee
defining $\;\mu =\mu(n_B)\;$. At zero baryon density $\;n_B=0\;$, as is clear 
from (62), one has $\;\mu=0\;$.

When the chemical potential is zero, the density of  quarks becomes
\be
\rho_q =\zeta_q \frac{3\Theta^3}{4\pi^2}\zeta(3) \qquad (\mu=0) ,
\ee
where $\;\zeta(3)=1.20206\;$. For gluons, the chemical potential is always 
zero, so their density is
\be
\rho_g =\zeta_g\frac{\Theta^3}{\pi^2}\zeta(3) .
\ee

Finally, for the specific heat of the Stefan--Boltzmann plasma we find
\be
C_{SB} \equiv \frac{\partial \ep_{SB}}{\partial\Theta} =
\frac{2\pi^2}{15}\left ( \zeta_g +\frac{7}{4}\zeta_q\right )\Theta^3 +
\frac{1}{2}\zeta_q\mu^2\Theta^2\frac{\mu^2-\pi^2\Theta^2}{3\mu^2+\pi^2\Theta^2}.
\ee

The results of numerical calculations [123-125] for the mixed system, 
consisting of the quark--gluon plasma with the spectrum (51) and of hadrons 
in the excluded--volume approximation, will be presented below for three 
different situations.

\subsection{SU(2) Quarkless System}

The system consists of unbound gluons and of glueballs that are bound gluon 
clusters. The experimental status of glueballs is yet uncertain, though there 
are suggestions [126,127] to interpret a narrow resonance appearing in 
proton--proton collisions as a scalar glueball. Pure gluodynamics is 
often studied because it is easier, than the full chromodynamics, for Monte
Carlo lattice simulations.

Glueball masses have been computed in the lattice gauge theory for both 
$\;SU(2)\;$ [43,46] as well as for $\;SU(3)\;$ [128-130] cases. The lattice 
results are close to the bag--model calculations [131-133]. Here we accept 
the glueball masses found in the bag--model approach [132,133]. The 
corresponding glueball characteristics are given in Table 2. The radius 
$\;r_2\;$ of the lightest glueball with $\;m_2=960\;MeV\;$ is a fitting 
parameter which is taken as $\;r_2=1.2\;fm\;$. The constant $\;B\;$ in (51)
is chosen to be $\;B=(165\;MeV)^4\;$. The gluon degeneracy factor is 
$\;\zeta_g=6\;$ for the $\;SU(2)\;$ case.

The results of our calculations are displayed in Figs.14-18, where the 
glueball probability $\;w_G\;$ and the gluon probability$\;w_g\;$ are defined 
by
$$ w_G\equiv \sum_i^{glueballs}z_i\frac{\rho_i}{\rho} , \qquad
w_g \equiv \frac{\rho_g}{\rho} = 1 - w_G . $$
The relative energy density is compared with the lattice data [134,135]. 
The reference Stefan--Boltzmann plasma here corresponds also to the 
quarkless case, $\;N_f=0\;$. Deconfinement occurs at $\;\Theta_d=215\;MeV\;$ 
as a second--order transition, which is in agreement with the lattice 
simulations.

\subsection{SU(3) Quarkless System}

The glueball parameters are taken from Table 2. The constant $\;B\;$ in 
spectra (51) is $\;B=(235\;MeV)^4\;$. The gluon degeneracy factor for the 
$\;SU(3)\;$ case is $\;\zeta_g=16\;$.

Varying the radius $\;r_2\;$ of the lightest glueball, we have three 
possibilities: (i) $\;r_2 < r_c\;$, where $\;r_c=0.8fm\;$; then deconfinement 
is a gradual crossover. (ii) $\;r_2=r_c\;$; in this case deconfinement is a 
$\;2\;$--order transition. (iii) $\;r_2 > r_c\;$; then $\;1\;$--order 
transition occurs. These possibilities are illustrated in Figs.19-23, where 
the entropy density at $\;\mu_i=0\;$ is $\;s=\bt(\ep+p)\;$ and the reference 
Stefan--Boltzmann entropy density is 
$\;s_{SB} =\bt(\ep_{SB} + p_{SB})=4\bt p_{SB}\;$. The relative entropy 
density is compared with the lattice numerical simulations [136]. The latter 
agrees with our results if $\;r_2 > r_c\;$, so that deconfinement becomes a 
$\;1\;$--order transition at about $\;\Theta_d=230\;MeV\;$.

Emphasize the importance of taking into consideration glueball interactions: 
When these are absent, that is $\;r_2=0\;$, the behaviour of the system 
is unphysical.

\subsection{SU(3) System with Quarks}

The constituents of the system are taken as follows. Consider quarks of two 
flavours, $\;q=\{ u,d\}\;$, and the corresponding antiquarks 
$\;\bar q =\{\bar u,\bar d\}\;$ with the masses $\;m_q=m_{\bar q}=7\;MeV\;$. 
The degeneracy factor for each pair of up and down quarks is 
$\;\zeta_q =2\times N_f\times N_c =12\;$, and the same for antiquarks, 
$\;\zeta_{\bar q}=12\;$. Gluons have the degeneracy factor 
$\;\zeta_g=2\times (N_c^2 -1) =16\;$. From the long list of the known 
hadrons, we include only those with the lightest masses, which mainly 
contribute to thermodynamics. These are unflavoured mesons (Table 3), strange 
mesons (Table 4) and light baryons (Table 5).

For the radius of the lightest hadron, that is of a pion, we take 
$\;r_2=r_\pi=0.56\;fm\;$. The radii of all other clusters are expresses 
through $\;r_\pi\;$ using (45). For the mean--field parameter in (51), we 
accept $\;B^{1/4}=210\;MeV\;$. The results of calculations are shown in 
Figs.24-26, where the hadron cluster probability is defined as
$$ w_c\equiv \sum_i^{clusters} z_i\frac{\rho_i}{\rho} . $$
Deconfinement is found to be rather a continuous crossover--like transition 
at $\;\Theta_d=166\;MeV\;$, which is close to lattice data [137].

\section{Thermodynamic Restriction Rule}

Invoking this or that approximation, one gets an {effective thermodynamic 
potential}, for instance, an effective grand potential 
$\;\Om=\Om(\Theta,V,\mu,\vp)\;$, involving some auxiliary functions 
$\;\vp=\{\vp_j\}\;$ depending on thermodynamics parameters, temperature 
$\;\Theta\;$, volume $\;V\;$, and a set $\;\mu=\{\mu_i\}\;$ of chemical 
potentials. Thus, effective spectra in (51) contain the mean field 
$\;\vp\equiv B/\rho\;$. In the excluded--volume approximation, the free 
volume of the system is factored with the quantity 
$\;\xi=1-\sum_i\rho_iv_i\;$, as is seen from (42). Both $\;\rho\;$ and 
$\;\rho_i\;$ are functions of $\;\Theta,V,\mu\;$. In the cut--off model of 
Section 5, the effective cut--off momentum $\;k_0\;$ is a function of 
$\;\Theta\;$.

Effective thermodynamic potentials are to be handled with great caution. 
Really, if one calculates, e.g., the pressure
\be
p=-\frac{\prt\Om}{\prt V} = -\frac{\Om}{V} =
\frac{\Theta}{V}\ln{\rm Tr}\;e^{-\bt H}
\ee
in two different ways, as the derivative $\;(-\prt\Om/\prt V)\;$ or as the 
ratio $\;(-\Om/V)\;$, then the answers can be different when $\;\Om\;$ 
includes auxiliary functions depending on $\;V\;$. This would mean that the 
relation (66) breaks. The same concerns the energy density, entropy density, 
and the cluster densities, respectively,
$$ \ep=\Theta\frac{\prt p}{\prt\Theta} - p + \sum_i\mu_i\rho_i =
\frac{1}{V}\lgl\hat E\rgl , $$
$$ s =\frac{\prt p}{\prt\Theta} =
\bt\left (\ep + p - \sum_i\mu_i\rho_i\right ) , $$
\be
\rho_i =\frac{\prt p}{\mu_i} =\frac{1}{V}\lgl\hat N_i\rgl ,
\ee
which may be defined in two ways, as first derivatives of pressure or as 
the corresponding statistical averages. Definition (67) can also become 
broken when auxiliary functions depend on thermodynamic variables. This 
kind of inconsistency occurs as well for the second derivatives, such as 
the specific heat
\be
C_V =\frac{\prt\ep}{\prt\Theta} =\frac{\bt^2}{V}\left ( \lgl\hat E^2\rgl -
\lgl\hat E\rgl^2\right )
\ee
or the isothermic compressibility
\be
\kappa_T=-\frac{1}{V}\left (\frac{\prt p}{\prt V}\right )^{-1} =
\frac{\bt}{\rho^2V}\left ( \lgl\hat N^2\rgl - \lgl\hat N\rgl^2\right ) .
\ee

These inconsistencies in defining thermodynamic characteristics in two ways, 
thermodynamic and statistical, of course, are not pleasant. Moreover, the 
difference between these two ways is not only quantitative, but can also 
become drastic, especially for systems with phase transitions. It is even 
possible to give examples when the definition through the derivatives yields 
unphysical divergencies in the energy and entropy densities at the phase 
transition point. This, for instance, happens, as is easy to check, 
for a pure gluon model in the effective spectrum approximation.

The simplest procedure for avoiding the described troubles can be formulated 
as follows. Let $\;x\;$ be any of the thermodynamic variables $\;\Theta,V\;$ or
$\;\mu\;$. If $\;\Om\;$ is an effective grand potential including auxiliary 
functions depending on these thermodynamic variables, then
$$ \frac{\prt\Om}{\prt x} =\left ( \frac{\prt\Om}{\prt x}\right )_\vp +
\frac{\prt\Om}{\prt\vp}\cdot\frac{\prt\vp}{\prt x} . $$
It is just the second term here which causes all unpleasant problems. So, 
the decision is evident: the derivatives $\;\prt\Om/\prt x\;$ are to be 
understood in the restricted sense as
\be
\frac{\prt\Om}{\prt x} \ra \left ( \frac{\prt\Om}{\prt x}\right )_\vp .
\ee
The consent (70) may be called the {\it thermodynamic restriction rule}. We 
always employ this rule dealing with effective thermodynamic potentials. If 
the derivatives in (66)-(69) are understood in the sense of (70), then both 
ways of calculating thermodynamic characteristics yield the same answers.

Although with the restriction rule (70) we avoid the appearance of spurious 
terms so that all relations (66)-(69) become self--consistent, another problem
can arise when dealing with effective thermodynamic potentials. This is the 
occurrence of instability regions around a transition point, where either the 
specific heat $\;C_V\;$ or the isothermic compressibility $\;\kappa_T\;$ are 
negative. Below we illustrate this for the $\;SU(3)\;$ quarkless system of 
subsection 12.2 with the mean--field parameter $\;B=(225\;MeV)^4\;$. The 
results are shown in Figs.27-34. The unstable solutions appearing in the 
vicinity of transition points are related to the loss of convexity of 
pressure. To restore the convexity, we may resort to the Maxwell construction 
smoothing the corresponding thermodynamic potential [138]. Such a smoothing 
is shown in Figs.31 and 32. The behaviour of the resulting pressure and energy
density is in a reasonable agreement with lattice simulations [50]. 
Nevertheless, a slight dissatisfaction rests with the fact that instability 
regions occur not only around a $\;1\;$--order transition, where this would 
be more or less natural, but also in the vicinity of a continuous transition.

Instead of relying on the restriction rule (70), it would seem rational to 
define an effective thermodynamic potential, from the beginning, in such a 
way that all necessary thermodynamic relations would be automatically valid. 
This goal can be achieved [139] by redefining thermodynamic characteristics 
with the help of the shifts of the chemical potentials, 
$\;\mu_i\ra\mu_i-u_i\;$, pressure, $\;p\ra p+p'\;$, energy density, 
$\;\ep\ra\ep +\ep'\;$, and entropy density, $\;s\ra s+s'\;$ requiring that 
the shifting functions $\;u_i,p',\ep'\;$, and $\;s'\;$ guarantee the validity 
of (66) and (67). The latter then are named the self--consistency conditions 
[139]. In this case, (66) and (67) is a system of nonlinear differential 
equations, for the functions $\;u_i,p',\ep'\;$ and $\;s'\;$, in partial 
derivatives with respect to the variables $\;\Theta,V\;$ and $\;\mu_i\;$. 
Such a system has no unique solution, especially when boundary conditions 
are not known. To extract a solution from the self--consistency 
equations needs several additional heuristic assumptions and fitting 
parameters. Some simplification comes from the guideline prescribed by 
mean--field approximations [140-142].

\section{Principle of Statistical Correctness}

In this section we present a new principle allowing a correct construction of 
effective thermodynamic potentials. This principle, as compared to the 
self--consistency conditions, is: (i) more general, yielding these conditions 
but not conversely; (ii) much simpler to deal with; (iii) unambiguous, 
providing a unique solution.

Let an effective thermodynamic potential $\;\Om_{eff}=\Om_{eff}(\vp)\;$ 
include a set $\;\vp=\{\vp_i(x)\}\;$ of auxiliary functions depending on 
arbitrary variables $\;x\;$. The latter, in particular, may incorporate space 
and thermodynamic variables. First of all, it is necessary to understand that 
not any effective potential can have sense, however reasonable it may look. 
Each thermodynamic potential, to be accepted as such, must have the properties
formulated below.

\vspace{5mm}

{\it Property 1}. {\bf Statistical Representability}:

\vspace{2mm}

An effective thermodynamic potential $\;\Om_{eff}\;$ represents an equilibrium
statistical system if and only if it has the Gibbs form
\be
\Om_{eff}(\vp) =\Om\left [ H_{eff}(\vp)\right ] ,
\ee
where
\be
\Om[H] \equiv -\Theta\ln{\rm Tr}\;e^{-\bt H} ,
\ee
depending on auxiliary functions only through an effective Hamiltonian 
$\;H_{eff}= H_{eff}(\vp)\;$. Such a thermodynamic potential is called 
{\it statistically representable}.

\vspace{5mm}

In this way, if one invents an effective thermodynamic potential, even 
pronouncing seemingly plausible words, this does not mean that the invented 
potential describes some statistical system. If the potential is not 
statistically representable it represents no equilibrium statistical system. 
For example, a thermodynamic potential in the excluded--volume approximation 
is not statistically representable. Although the latter approximation may 
occasionally give a reasonable description, but in general it is not 
trustworthy. The excluded--volume approximation may be used, because of its 
simplicity, as a first attempt of understanding the qualitative behaviour of 
a system, but it should be always followed by a more reliable approximation.

\vspace{5mm}

{\it Property 2}. {\bf Thermodynamic Equivalence}:

\vspace{2mm}

A statistical system described by a given Hamiltonian $\;H_{giv}\;$ is
thermodynamically equivalent to a system modeled by an effective Hamiltonian 
$\;H_{eff}\;$ if and only if their thermodynamic potentials are statistically 
representable,
\be
\Om_{giv} =\Om[H_{giv}], \qquad \Om_{eff} =\Om[H_{eff}] ,
\ee
and are equal to each other,
\be
\Om[H_{giv}] =\Om[H_{eff}] .
\ee
The corresponding Hamiltonians are called {\it thermodynamically equivalent}.

For the case of infinite matter, such as nuclear matter, the equality (74) 
can be softened by requiring the validity of the asymptotic, in the 
thermodynamic limit, equality
$$ \lim_{V\ra\infty}\frac{1}{V}
\left ( \Om[H_{giv}] - \Om[H_{eff}]\right ) = 0. $$

\vspace{5mm}

{\it Property 3}. {\bf Statistical Equilibrium}:

\vspace{2mm}

The necessary condition for an equilibrium statistical system modelled by an 
effective Hamiltonian $\;H_{eff}(\vp)\;$ to be thermodynamically equivalent 
to a given statistical system with $\;H_{giv}\;$ is the equilibrium condition
\be
\left \lgl\frac{\delta}{\delta\vp}H_{eff}(\vp)\right\rgl = 0 ,
\ee
where the variation over $\;\vp\;$ implies a set of variations with respect 
to each $\;\vp_i\;$ and
$$ \lgl\hat A\rgl \equiv
\frac{{\rm Tr}\hat A\exp (-\bt H_{eff})}{{\rm Tr}\exp(-\bt H_{eff})} . $$

The proof of (75) is straightforward basing on the statistical 
representability (71), thermodynamic equivalence (74) and the fact that 
$\;\Om_{giv}\;$does not depend on $\;\vp\;$.

Now we can formulate the central notion: 

\vspace{3mm}

{\bf Principle of Statistical Correctness}:

\vspace{2mm}

{\it An effective thermodynamic potential is statistically correct if it is 
statistically representable with an effective Hamiltonian satisfying the  
condition (75) of statistical equilibrium}.

\vspace{2mm}

As is evident, the self--consistency conditions for the first-order 
derivatives (66) and (67) immediately follow from (75). Moreover, the 
self--consistency conditions for the second--order derivatives (68) and 
(69) also follow from (75) as well as such conditions for the derivatives 
of arbitrary order. While if one finds the shifting functions from the 
first--order self--consistency conditions (66) and (67), the second--order 
conditions (68) and (69) are not necessarily fulfilled.

We shall also say that an effective Hamiltonian is statistically correct 
if it satisfies (75). The same can be said about an approximation leading 
to a statistically correct Hamiltonian. For instance, the correlated 
mean--field approximation of Section 9, involving conditions (26) or (27), 
or (31), is statistically correct.

\section{Clustering Quark--Hadron Matter}

To obtain a statistically correct description of the quark--gluon plasma 
clustering into hadron states, let us use the correlated mean--field 
approximation [35] leading to the Hamiltonian
$$ H =\sum_i H_i +CV , $$
$$ H_i=\sum_k\om_i(k)a_i^\dgr(\sk)a_i(\sk) , $$
\be
\om_i(k) =\sqrt{k^2+m_i^2}+U_i-\mu_i
\ee
discussed in Section 9.

Consider the case of the conserved baryon number $\;N_B=\sum_iN_i^B\;$ with 
$\;N_i^B\equiv B_iN_i\;$, where $\;B_i\;$ is the baryon number of an 
$\;i\;$--cluster. For an equilibrium system, the relation
\be
\mu_i =B_i\mu_B
\ee
holds between the chemical potential $\;\mu_i\;$ and the baryon potential 
$\;\mu_B\;$. The latter may be defined as a function $\;\mu_B(n_B)\;$ of the 
baryon density
\be
n_B\equiv\frac{N_B}{V} =\sum_iB_i\rho_i .
\ee

The index $\;i\;$ enumerates the constituents. The total set $\;\{ i\}\;$ of 
these indices consists of two different groups, 
$\;\{ i\}=\{ i\}_{pl}\bigcup\;\{ i\}_{cl} \;$. The first group 
$\;\{ i\}_{pl}\;$ corresponds to the plasma constituents, quarks, 
antiquarks, and gluons, which are elementary particles, thus, having the 
compositeness number $\;z_i=1\;$. The second group $\;\{ i\}_{cl}\;$ 
enumerates hadron clusters that are bound states with compositeness 
numbers $\;z_i\geq 2\;$. Respectively, the total density of matter
\be
\rho =\sum_iz_i\rho_i =\rho_{pl} +\rho_{cl} ,
\ee
consists of two terms
$$ \rho_{pl} =\sum_{\{ i\}_{pl}}\rho_i , \qquad
\rho_{cl}=\sum_{\{ i\}_{cl}}z_i\rho_i $$
corresponding to the plasma density $\;\rho_{pl}\;$ and the cluster density 
$\;\rho_{cl}\;$.

Define the plasma mean fields $\;U_i\;$, when $\;i\in\{ i\}_{pl}\;$, as
\be
U_i =U(\rho)=\rho\int V(r)s(r)d\sr ,
\ee
where $\;V(r)\;$ is a confining potential and $\;s(r)\;$, screening function. 
Before substituting into (80) a concrete confining potential, let us 
emphasize the general properties which the plasma mean field $\;U(\rho)\;$ 
must satisfy to. These properties are
$$ U(\rho)\ra\infty \qquad (\rho\ra 0) , $$
\be
U(\rho)\ra 0 \qquad (\rho\ra\infty) .
\ee
The upper line in (81) means that quarks and gluons cannot exist as unbound 
particles at low density, that is, the colour confinement must occur as 
$\;\rho\ra 0\;$. Or one may say that quarks and gluons cannot exist as free 
particles outside dense nuclear matter. The lower line in (81) reflects the 
phenomenon of asymptotic freedom.

There are different types of confining potentials, linear, quadratic, 
logarithmic, and with noninteger powers. For example, the interaction between 
a heavy quark and its antiquark is usually taken in the form of the Cornell 
potential [143,144] with the linear confining term. This form of the potential
is confirmed by $\;QCD\;$ calculations [145] and by lattice simulations [146].
The quadratic confining potential is also quite popular [77,80,91]. In 
addition, the intensity of mutual interactions between three plasma 
constituents, quarks, antiquarks, and gluons, is different [145,147]. The 
confining potential $\;V(r)\;$ in (80) is assumed to be an averaged potential 
of the form
\be
V(r) = Ar^\nu \qquad (0\leq \nu\leq 2) .
\ee

The screening function $\;s(r)=c(r/a)\;$ is usually [91] scaled with the mean 
interparticle distance $\;a\equiv\rho^{-1/3}\;$. Therefore, the plasma mean 
field (80) with the confining potential (82) can be written as
\be
U(\rho) =J^{1+\nu}\rho^{-1/3} ,
\ee
where
$$ J^{1+\nu}\equiv 4\pi A\int_0^\infty c(x)x^{2+\nu}dx . $$
We can calculate the constant $\;J\;$ if $\;A\;$ and $\;c(x)\;$ are known. 
Alternatively, we may treat $\;J\;$ as a free parameter. The value of $\;J\;$ 
can be estimated as follows. Accept that at the normal quark density 
$\;\rho_0\;$ the plasma mean field (83) becomes
\be
U(\rho_0) = 3E_0 = 3\rho_0^{1/3} ,
\ee
where the factor $\;3\;$ stands for the three plasma constituents. Then 
from (83) and (84) we obtain
\be
J = 3^{1/(1+\nu)}\rho_0^{1/3} .
\ee
Thus, for the linear confinement, $\;\nu=1\;$, we get $\;J=272\;MeV\;$, while
for the quadratic confinement, $\;\nu=2\;$, we have $\;J=226\;MeV\;$.

For the mean field of an $\;i\;$-cluster we may write
\be
U_i =\sum_{\{ j\}_{cl}}\Phi_{ij}\rho_j + 
z_i\left [ U(\rho) - U(\rho_{cl})\right ] ,
\ee
where the first term describes the interaction of the given cluster with 
other clusters, and the second term corresponds to the interaction of this 
cluster with the quark--gluon plasma. The interaction potentials between 
clusters can be scaled according to (20), which permits to express the 
interaction integrals (29) through one scaling integral as
\be
\Phi_{ij} =z_iz_j\Phi .
\ee
Taking into account (83) and (87) reduces (86) to
\be
U_i=z_i\Phi\rho_{cl} +
z_iJ^{1+\nu}\left (\rho^{-\nu/3} -\rho_{cl}^{-\nu/3}\right ) .
\ee

In the effective Hamiltonian (76) with the mean fields (83) and (88) the 
role of auxiliary functions is played by the densities $\;\rho\;$ and 
$\;\rho_{cl}\;$. So, for the equilibrium conditions (75) we have
\be
\left\lgl\frac{\prt H}{\prt\rho}\right\rgl =0 , \qquad
\left\lgl\frac{\prt H}{\prt\rho_{cl}}\right\rgl = 0 .
\ee
From (89) we have two variational equations of the type (31), whose solution, 
up to a constant, is easy to find:
\be
C =\frac{\nu}{3-\nu} J^{1+\nu}
\left ( \rho^{1-\nu/3} -\rho_{cl}^{1-\nu/3}\right ) -
\frac{1}{2}\Phi\rho_{cl}^2 .
\ee
In this way, the effective Hamiltonian is completely defined and we can pass 
to particular calculations.

\subsection{Pure Gluon System}

Imagine an extreme situation when only gluons can exist. This case may be 
obtained from the general model by putting all degeneracy factors zero except 
that of gluons, $\;\zeta_g\neq 0\;$. Then $\;\rho=\rho_g\;$ and 
$\;\rho_{cl}=0\;$. Employing the Boltzmann approximation, we find [148] that 
there occurs a first order transition vacuum--gluon plasma at
$$ \Theta_d = J\left [ \frac{\nu}{3-\nu}\exp\left ( 1 -\frac{\nu}{3}\right )
\left (\frac{\pi^2}{\zeta_g}\right )^{\nu/3}\right ]^{1/(1+\nu)} . $$
Below $\;\Theta_d\;$ there is exactly vacuum, empty space, with zero energy 
density $\;\ep=0\;$. Gluons appear at $\;\Theta_d\;$ with a jump. The relative
latent heat at $\;\Theta_d\;$ is
$$ \frac{\Delta\ep_d}{\ep_{SB}}=
\frac{1+\nu}{\nu}\exp\left ( 1 -\frac{3}{\nu}\right ) . $$
The degeneracy of gluons for the $\;SU(3)\;$ case is $\;\zeta=16\;$. 
For the linear comfinement with $\;\nu=1\;$, we get $\;\Theta_d=248\;MeV\;$ 
and $\;\Delta\ep_d/\ep_{SB}=0.27\;$. For the harmonic confinement, when 
$\;\nu=2\;$, the vacuum--gluon transition happens at a higher temperature 
$\;\Theta_d=285\;MeV\;$ and the latent heat is larger, 
$\;\Delta\ep_d/\ep_{SB}=0.91\;$.

\subsection{SU(2) Gluon--Glueball System}

The gluon degeneracy factor for the $\;SU(2)\;$ case is $\;\zeta=6\;$. As 
the scaling integral we take $\;\Phi=\Phi_{22}/4\;$. So that the interactions 
between glueballs are found from (87). We consider the glueballs listed in 
Table 2. There are two fitting parameters for which we accept $\;J=175\;MeV\;$ 
and $\;\Phi_{22}=38.42\;GeV\; fm^3\;$, so that $\;\Phi=9.61\;GeV\; fm^3\;$. 
The results of our calculations [148-151] are shown in Figs.35-37 for the  
quadratic confinement, $\;\nu\approx 2\;$. Actually, the results do not 
change much in the interval $\;1.5\leq\nu\leq 2\;$. We have mainly used 
$\;\nu=1.86\;$. Deconfinement is a second--order transition at 
$\;\Theta_d=210\;MeV\;$. As is seen, the agreement with the lattice 
simulations [135] is beautiful.

\subsection{SU(3) Gluon--Glueball Mixture}

For the gluon degeneracy factor we have $\;\zeta_g=16\;$. The power $\;\nu\;$ 
of the confining interaction is, as in the previous subsection, close to
quadratic. But the fitting parameters are $\;J=225\;MeV\;$ and 
$\;\Phi=3.84\;GeV\; fm^3\;$. The glueball characteristics are again taken from 
the Table 2. Calculations show [148,149] that a first--order transition occurs
at $\;\Theta_d=225\;MeV\;$ with the relative latent heat 
$\;\Delta\ep_d/\ep_{SB}=0.23\;$. The agreement with the Monte Carlo lattice 
simulations [50,51,152] is also very good.

\subsection{Mixed Quark--Gluon--Meson System}

Consider the case of zero baryon density, $\;n_B=0\;$. Take the quarks of 
two flavours, $\;u\;$ and $\;d\;$, and the related antiquarks, $\;\bar u\;$ 
and $\;\bar d\;$. Assume, for simplicity, that all their masses are equal, 
$\;m_u=m_d=7\;MeV\;$. The degeneracy factor for each kind of quarks is 
$\;\zeta_u=\zeta_d=2N_c=6\;$. This factor for gluons is $\;\zeta_g=16\;$. 
Hadrons are represented by mesons from Tables 3 and 4. Following Section 11, 
we accept $\;\Phi_{33}=315\;MeV\;fm^3\;$, so that 
$\;\Phi=\Phi_{33}/9=35\;MeV\;fm^3\;$.
The plasma interaction parameter $\;J=225\;MeV\;$ is the same as in the 
previous subsection, as well as $\;\nu\approx 2\;$ corresponding to quadratic 
confinement. So, here we do not add any new fitting parameters.

Our calculations [148,149] displayed in Fig.40 prove that there is no sharp 
phase transition but there is a gradual crossover. The deconfinement 
transition can be attributed to the temperature $\;\Theta_d=150\;MeV\;$ where 
the relative specific heat $\;C_V/C_{SB}\;$ has a maximum. The latter is 
finite and look rather as a Schottky anomaly [153] than as a narrow divergent 
peak typical of a second--order phase transition. The agreement of our 
results with the lattice--simulation data [137] is again quite good. The 
lattice results [137] indicate that the deconfinement transition is really 
continuous. From the point of view of $\;QCD\;$ this can be understood as 
follows. The role of the quark term in the $\;QCD\;$ Lagrangian is similar 
to that of an external magnetic field applied to a spin system. In the 
presence of a magnetic field, the ferromagnet--paramagnet transition in 
simple spin systems becomes a continuous crossover.

\subsection{Finite Baryon Density}

Here we extend the consideration to nonzero baryon density. The parameters 
$\;J\;$ and $\;\Phi\;$ are the same as in the previous subsection. Again, 
we study the two--flavour case with the same characteristics. We take mesons 
from Table 3, protons and neutrons from Table 5, and multiquarks from Table 1.
For a six--quark cluster we accept $\;m_6=1944\;MeV\;$. The Bose--Einstein 
condensate of six--quarks occurs when $\;\om_6(0)=0\;$. Then, the baryon 
potential is
$$ \mu_B =\frac{1}{2}m_6 + 3\Phi\rho_{cl} + 3J^{1+\nu}
\left ( \rho^{-\nu/3} + \rho_{cl}^{-\nu/3}\right ) , $$
and the density of six--quarks consists of two terms:
$$ \rho_6 =\frac{\zeta_6}{(2\pi)^3}\int n_6(k)d\sk + \rho_6^0\; . $$

We have analysed the behaviour of several probabilities [154-157] as functions
of temperature $\;\Theta\;$ and relative baryon density $\;n_B/n_{0B}\;$. This
is demonstrated in Fig.41 for the plasma probability
$$ w_{pl} =\frac{1}{\rho}\left ( \rho_g + \rho_u + \rho_{\bar u} +
\rho_d +\rho_{\bar d}\right ) , $$
in Fig.42 for the pion probability
$$ w_\pi =\frac{2}{\rho}
\left (\rho_{\pi^+} +\rho_{\pi^-} +\rho_{\pi^0}\right ) , $$
in Fig.43 for a summarized, excluding pions, probability of other mesons
$$ w_{\eta\rho\om} =\frac{2}{\rho}\left (\rho_\eta + \rho_{\rho^+} +
\rho_{\rho^-} +\rho_{\rho^0} + \rho_\om\right ) \equiv w_{mes} , $$
in Fig.44 for the nucleon probability
$$ w_3 =\frac{3}{\rho}\left ( \rho_p + \rho_{\bar p} + \rho_n +
\rho_{\bar n}\right ) , $$
in Fig.45 for the six--quark probability
$$ w_6 =\frac{6}{\rho}\left ( \rho_6 + \rho_{\bar 6}\right ) , $$
and in Fig. 46 for the probability of condensed six--quark clusters
$$ w_6^0 =\frac{6}{\rho}\rho_6^0 \; . $$

In addition, we present here some other thermodynamic characteristics 
permitting to better understand the features of the deconfinement transition. 
The ratio of pressure over energy density, which has the meaning of the 
effective sound velocity squared,
$$ c^2_{eff} =\frac{p}{\ep} \; , $$
is given in Fig. 47. At $\;n_B=0\;$ the temperature dependence of 
$\;c_{ef}^2\;$ agrees with that reconstructed from the lattice data [158]. 
The reduced specific heat
$$ \sigma_V = \frac{\Theta}{\ep}\frac{\prt\ep}{\prt\Theta} $$
and the dimensionless compressibility coefficient
$$ \kappa_T = \left (\frac{n_B}{J^4}\frac{\prt p}{\prt n_B}\right )^{-1} $$
are depicted in Figs. 48 and 49, respectively. The transition line can be 
ascribed to the maximum of the inverse compressibility coefficient 
$\;\kappa_T^{-1}\;$ which can be called [111] the compression modulus 
(see Fig.50).

We shall not discuss in detail the peculiarities of the calculated 
thermodynamic functions. This is because, as we think, the presented figures 
already give a good visual demonstration, and also in order not to make this 
review too long. Let us only emphasize that the deconfinement transition 
is a continuous crossover becoming smoother and smoother with increasing 
baryon density. The deconfinement at a fixed low temperature and rising 
baryon density is due to the disintegration of hadrons into unbound quarks. 
When both temperature and baryon density increase, the deconfinement is a 
result of the hadron disintegration as well as of the generation from 
vacuum of quarks and gluons.

Concluding we may state that taking into account the coexistence of hadrons 
and of the quark--gluon plasma is vitally important for constructing a unified
approach being in agreement with the lattice--simulation data. The gradual 
character of the deconfinement transition, occurring through a mixed
hadron--plasma state, rules out those predictions that have been based on 
a sharp first--order phase transition. This concerns the interpretation 
of signals of the quark--gluon plasma at heavy ion collisions [8-14,159] and 
the hadronization scenario related to the evolution of early universe after 
the Big Bang [4]. The quantitative predictions of our approach can be improved
in several ways, for instance, by including more kinds of particles or 
by invoking more elaborate interaction potentials [160]. However, we do 
hope that qualitatively the picture will remain the same.

\vspace{5mm}

{\bf Acknowledgements}

\vspace{2mm}

We are greatly indebted to A.M.Baldin for sparking one of the authors (V.I.Y.)
by the problem considered, permanent support and many valuable advises. We
acknowledge, with pleasure and gratitude, a number of fruitful discussions 
with our colleagues from the Relativistic Nuclear Physics Group at Dubna, 
especially to V.V.Burov, V.K.Lukyanov, and A.I.Titov. We appreciate the 
help of our coauthors collaborating with us at different stages of our work.

\newpage

{\bf Figure Captions}

\vspace{5mm}

{\bf Fig.1.}\\ 
Illustration of a cluster fusion.

\vspace{5mm}

{\bf Fig.2.}\\ 
$\;6q\;$--probability vs. the mass of a $\;6q\;$--cluster at $\;\Theta=0\;$ 
and $\;\rho=\rho_0\;$.

\vspace{5mm}

{\bf Fig.3.}\\
Cluster probabilities as functions of the  relative density at 
$\;\Theta=200\;MeV\;$.

\vspace{5mm}

{\bf Fig.4.}\\
Nucleon probability as a function of the relative density at 
different temperatures.

\vspace{5mm}

{\bf Fig.5.}\\
$\;6q\;$--probability vs. relative density at $\;\Theta=0\;$.

\vspace{5mm}

{\bf Fig.6.}\\
Nucleon, $\;6q\;$--cluster, and unbound quark probabilities as functions of 
the  relative density at $\;\Theta=0\;$.

\vspace{5mm}

{\bf Fig.7.}\\
Probabilities of unbound quarks and of bound clusters vs. relative density 
at $\;\Theta=30\;MeV\;$.

\vspace{5mm}

{\bf Fig.8.}\\
Cluster probabilities vs. relative density at $\;\Theta=50\;MeV\;$.

\vspace{5mm}

{\bf Fig.9.}\\
Cluster probabilities at $\;\Theta=100\;MeV\;$.

\vspace{5mm}

{\bf Fig.10.}\\
Cluster probabilities at $\;\Theta=200\;MeV\;$. Dashed lines show the points 
of first order phase transitions. Between these points the matter is a 
stratified gas--liquid mixture.

\vspace{5mm}

{\bf Fig.11.}\\
Nucleon, $\;6q\;$--cluster, and quark probabilities as functions of 
temperature in $\;MeV\;$ at the normal density $\;\rho=\rho_0\;$.

\vspace{5mm}

{\bf Fig.12.}\\
Cluster probabilities vs. temperature at the fixed density $\;\rho=5\rho_0\;$.

\vspace{5mm}

{\bf Fig.13.}\\
Phase portrait for the baryon rich quark--hadron matter. Along the dashed 
line the compressibility is divergent.

\vspace{5mm}

{\bf Fig.14.}\\
Glueball probability for the $\;SU(2)\;$ quarkless system as a function of 
temperature.

\vspace{5mm}

{\bf Fig.15.}\\
Comparison of glueball and gluon probabilities for the $\;SU(2)\;$ quarkless 
system.

\vspace{5mm}

{\bf Fig.16.}\\
Relative energy density for the $\;SU(2)\;$ quarkless system (solid line) as 
compared with the lattice Monte Carlo data (Engels et al., 1981).

\vspace{5mm}

{\bf Fig.17.}\\
Relative energy density for the $\;SU(2)\;$ quarkless system (solid line) 
compared with the lattice numerical simulations (Engels et al., 1989).

\vspace{5mm}

{\bf Fig.18.}\\
Reduced specific heat for the $\;SU(2)\;$ gluon--glueball mixture. At the 
deconfinement temperature, specific heat diverges.

\vspace{5mm}

{\bf Fig.19.}\\
Glueball probability for the $\;SU(3)\;$ quarkless system at several values of
the lightest glueball radius: (1) $\;r_2=0\;$; (2) $\;r_2=0.5\;fm\;$; 
(3) $\;r_2=0.7\;fm\;$; (4) $\;r_2=0.8\;fm\;$.

\vspace{5mm}

{\bf Fig.20.}\\
Relative energy density for the $\;SU(3)\;$ quarkless system at: 
(1) $\;r_2=0\;$; (2) $\;r_2=0.5\;fm\;$; (3) $\;r_2=0.7\;fm\;$; 
(4) $\;r_2=0.8\;fm\;$; (5) $\;r_2=1\;fm\;$.

\vspace{5mm}

{\bf Fig.21.}\\
Relative entropy density for the $\;SU(3)\;$ quarkless system at 
$\;r_2=0.82\;fm\;$ compared with the lattice numerical data 
(Brown et al., 1988).

\vspace{5mm}

{\bf Fig.22.}\\
Reduced specific heat for the $\;SU(3)\;$ gluon--glueball mixture at 
$\;r_2=0.8\;fm\;$.

\vspace{5mm}

{\bf Fig.23.}\\
Relative energy density for the $\;SU(3)\;$ quarkless system at 
$\;B^{1/4}=210\;MeV\;$ and $\;r_2=0.6\;fm\;$.

\vspace{5mm}

{\bf Fig.24.}\\
Hadron cluster probability for the mixture of quark--gluon plasma and 
hadrons at zero baryon density.

\vspace{5mm}

{\bf Fig.25.}\\
Relative energy density for the mixture compared with the lattice 
numerical calculations (\c{C}elik et al., 1985).

\vspace{5mm}

{\bf Fig.26.}\\
Reduced specific heat for the mixture of quark--gluon plasma and hadrons at 
zero baryon density.

\vspace{5mm}

{\bf Fig.27.}\\
Total density as a function of temperature for the $\;SU(3)\;$ gluon--glueball
mixture at different values of the lightest glueball radius: 
(1) $\;r_2=0.6\;fm\;$; (2) $\;r_2=0.8\;fm\;$; (3) $\;r_2=1\;fm\;$.

\vspace{5mm}

{\bf Fig.28.}\\
Gluon probability vs. temperature for the values of $\;r_2\;$ as in Fig.27.

\vspace{5mm}

{\bf Fig.29.}\\
Pressure of the gluon--glueball mixture vs. temperature for the same values 
of $\;r_2\;$ as in Fig.27.

\vspace{5mm}

{\bf Fig.30.}\\
Relative energy of the mixture for the values of $\;r_2\;$ as in Fig.27.

\vspace{5mm}

{\bf Fig.31.}\\
Smoothing of pressure in the unstable crossover region at $\;r_2=0.7\;fm\;$.

\vspace{5mm}

{\bf Fig.32.}\\
Smoothing of pressure in the region of first--order phase transition at 
$\;r_2=1\;fm\;$.

\vspace{5mm}

{\bf Fig.33.}\\
Relative smoothed pressure for the gluon--glueball system at 
$\;r_2=0.9\;fm\;$ (solid line) as compared with the lattice simulation 
data (Engels, 1991).

\vspace{5mm}

{\bf Fig.34.}\\
Relative smoothed energy at $\;r_2=0.9\;fm\;$ (solid line) compared with the 
lattice data (Engels, 1991).

\vspace{5mm}

{\bf Fig.35.}\\
Gluon probability for the corrected $\;SU(2)\;$ quarkless model.

\vspace{5mm}

{\bf Fig.36.}\\
Relative energy and pressure for the corrected $\;SU(2)\;$ gluon--glueball 
model. Circles and squares are lattice simulation data (Engels, 1989).

\vspace{5mm}

{\bf Fig.37.}\\
Reduced specific heat for the corrected $\;SU(2)\;$ quarkless model.

\vspace{5mm}

{\bf Fig.38.}\\
Gluon probability for the corrected $\;SU(3)\;$ gluon--glueball model.

\vspace{5mm}

{\bf Fig.39.}\\
Relative energy and pressure for the corrected $\;SU(3)\;$ quarkless model 
compared with the lattice Monte Carlo calculations (Brown et al., 1988; 
Karsch, 1989; Engels, 1991; Petersson, 1991).

\vspace{5mm}

{\bf Fig.40.}\\
Relative energy and pressure for the mixed quark--gluon--meson system.

\vspace{5mm}

{\bf Fig.41.}\\
Quark--gluon plasma probability on the temperature--baryon density plane.

\vspace{5mm}

{\bf Fig.42.}\\
The probability of $\;\pi\;$--mesons.

\vspace{5mm}

{\bf Fig.43.}\\
The total probability of $\;\eta\;$--, $\;\rho\;$--, and $\;\omega\;$--mesons.

\vspace{5mm}

{\bf Fig.44.}\\
Nucleon probability.

\vspace{5mm}

{\bf Fig.45.}\\
The probability of six--quark clusters.

\vspace{5mm}

{\bf Fig.46.}\\
The probability of condensed six--quark clusters.

\vspace{5mm}

{\bf Fig.47.}\\
Pressure--to--energy ratio on the temperature--baryon density plane.

\vspace{5mm}

{\bf Fig.48.}\\
Reduced specific heat.

\vspace{5mm}

{\bf Fig.49.}\\
Dimensionless compressibility coefficient.

\vspace{5mm}

{\bf Fig.50.}\\
Dimensionless compression modulus as a function of temperature and baryon 
density.

\newpage

\begin{center}
{\bf Table Captions}
\end{center}

{\bf Table 1.} Multiquark parameters

\vspace{5mm}

{\bf Table 2.} Glueball parameters

\vspace{5mm}

{\bf Table 3.} Unflavoured meson parameters

\vspace{5mm}

{\bf Table 4.} Strange meson parameters

\vspace{5mm}

{\bf Table 5.} Baryon parameters

\newpage

\begin{center}

{\bf Table 1}

\vspace{5mm}

\begin{tabular}{|c|c|c|}\hline
$ mass $      & $ compositeness $ & $ degeneracy $ \\
$ m_i\;(MeV)$ & $ number\; z_i $  & $ factor\;\zeta_i $ \\ \hline
    939       &     3             &    4   \\
   1944       &     6             &    9 \\
   2163       &     6             &    3 \\
   3521       &     9             &    4  \\
   4932       &    12             &    1 \\ \hline
\end{tabular}

\vspace{3cm}

{\bf Table 2}

\vspace{5mm}

\begin{tabular}{|c|c|c|}\hline
$ mass $      & $ compositeness $ & $ degeneracy $ \\
$ m_i\;(MeV)$ & $ number\; z_i $  & $ factor\;\zeta_i $ \\ \hline
    960       &     2             &    6   \\
   1290       &     2             &    6 \\
   1590       &     2             &    6 \\
   1460       &     3             &   11  \\
   1800       &     3             &   39 \\ \hline
\end{tabular}

\vspace{3cm}

{\bf Table 3}

\vspace{5mm}

\begin{tabular}{|c|c|c|c|}\hline
$mesons$& $ mass $      & $ compositeness $ & $ degeneracy $ \\
        & $ m_i\;(MeV)$ & $ number\; z_i $  & $ factor\;\zeta_i $ \\ \hline
$\pi^+$ &     140       &     2             &    1   \\
$\pi^-$ &     140       &     2             &    1 \\
$\pi^0$ &     135       &     2             &    1 \\
$\eta$  &     548       &     2             &    1  \\
$\rho^+$&     770       &     2             &    3 \\ 
$\rho^-$&     770       &     2             &    3  \\
$\rho^0$&     770       &     2             &    3   \\
$\om$   &     782       &     2             &    3 \\ \hline
\end{tabular}

\newpage

{\bf Table 4}

\vspace{5mm}

\begin{tabular}{|c|c|c|c|}\hline
$mesons$    & $ mass $      & $ compositeness $ & $ degeneracy $ \\
            & $ m_i\;(MeV)$ & $ number\; z_i $  & $ factor\;\zeta_i $\\ \hline
$K^+$       &     494       &     2             &    1   \\
$K^-$       &     494       &     2             &    1 \\
$K^0$       &     498       &     2             &    2 \\
$\bar K^0$  &     498       &     2             &    2  \\ \hline
\end{tabular}

\vspace{3cm}

{\bf Table 5}

\vspace{5mm}

\begin{tabular}{|c|c|c|c|}\hline
$baryons$   & $ mass $      & $ compositeness $ & $ degeneracy $ \\
            & $ m_i\;(MeV)$ & $ number\; z_i $  & $ factor\;\zeta_i $\\ \hline
$ N $       &     939       &     3             &    4  \\
$\bar N$    &     939       &     3             &    4 \\
$\Delta$    &    1232       &     3             &   16 \\
$\bar\Delta$&    1232       &     3             &   16  \\ \hline
\end{tabular}

\end{center}

\end{document}